\UseRawInputEncoding
\documentclass[aps,showpacs,prd,twocolumn,nofootinbib,nobibnotes,]{revtex4-2}

\usepackage{graphicx}
\usepackage{amssymb}
\usepackage{amsmath}
\usepackage{color}
\usepackage{float}
\usepackage{ulem}
\usepackage{booktabs}
\usepackage{accents}
\usepackage{graphicx}
\usepackage{graphicx}
\usepackage{amsfonts}
\usepackage[colorlinks=true,
pdfstartview=FitV,linkcolor=blue,
citecolor=blue,urlcolor=blue,breaklinks=true]
{hyperref}
\usepackage[utf8]{inputenc}
\usepackage[T1]{fontenc}
\usepackage{array}
\usepackage{float}

\usepackage{booktabs}
\usepackage{multirow}

\usepackage{placeins}
\usepackage[dvipsnames]{xcolor}
\usepackage{csquotes}
\usepackage{bbold}
\usepackage{units}
\usepackage{tabularx}
\usepackage{enumitem}
\usepackage{subcaption}
\usepackage{dsfont}
\usepackage{upgreek}
\usepackage{ragged2e}
\usepackage{caption}
\newcolumntype{C}[1]{>{\centering\arraybackslash}m{#1}}
\renewcommand{\eqref}[1]{\mbox{Eq.~(\ref{#1})}}
\newcommand{\figref}[1]{\mbox{Fig.~(\ref{#1})}}

\definecolor{ForestGreen}{rgb}{0.13,0.55,0.13}
\usepackage{fixmath}
\newcommand{\orcid}[1]{\href{https://orcid.org/#1}{\includegraphics[width=10pt]{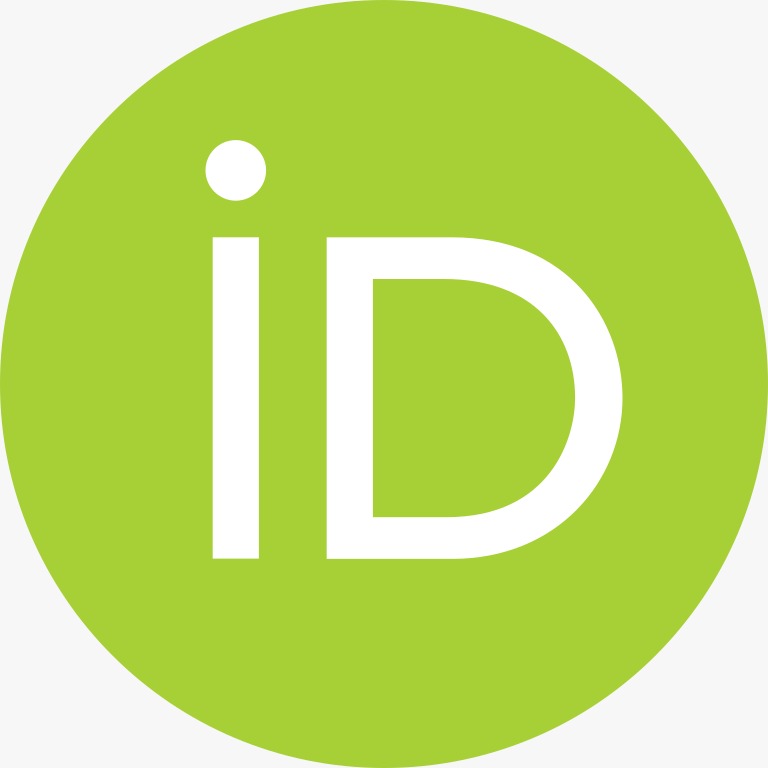}}}

\begin{document}

	\title{Multiplicity of surface polariton configurations in bi-isotropic media with anomalous Hall current}

\author{Alex Q. Costa\orcid{0009-0001-9812-7645}$^a$}
\email{costa.alex@discente.ufma.br, prof.costalex@gmail.com}
\author{Pedro D. S. Silva\orcid{0000-0001-6215-8186}$^{b}$}
\email{pedro.dss@ufma.br, pdiego.10@hotmail.com}
\author{Manoel M. Ferreira Jr.\orcid{0000-0002-4691-8090}$^{a, c}$}
\email{manojr.ufma@gmail.com, manoel.messias@ufma.br}
\affiliation{$^a$Programa de P\'{o}s-graduaç\~{a}o em F\'{i}sica, Universidade Federal do Maranh\~{a}o, Campus
	Universit\'{a}rio do Bacanga, S\~{a}o Lu\'is, Maranh\~ao, 65080-805, Brazil}
\affiliation{$^b$Coordena\c{c}\~ao do Curso de Ci\^encias Naturais - F\'isica, Universidade Federal do Maranh\~ao, Campus de Bacabal, Bacabal, Maranh\~ao, 65700-000, Brazil}
\affiliation{$^c$Coordenação do Curso de Física Bacharelado, Universidade Federal do Maranh\~{a}o, Campus Universit\'{a}rio do Bacanga, S\~{a}o Lu\'is, Maranh\~ao 65080-805, Brazil}

	\begin{abstract}

In this work, we investigate surface polaritons (SPs) arising at the interface between a simple dielectric and a bi-isotropic medium, in the lossless regime and in the presence of small losses. For the former,  non-Dyakonov surface polaritons can arise for positive permittivities and magnetoelectric parameters. General restrictions on the permittivities and bi-isotropic parameter (existence conditions) are derived. For the latter, the effect of small losses is considered, and the characteristic surface plasmon-polariton lengths are discussed. The propagation length can exhibit a local maximum for specific values of the magnetoelectric parameter, an unusual new behavior that can be regarded as a signature of these chiral surface polaritons. Surface waves in an interface involving a bi-isotropic medium with an anomalous Hall effect (AHE) are also examined. Multiple configurations for their occurrence are obtained, revealing a richer scenario with distinct regimes of propagation and coexistence in the parameter space. The presence of a non-null AHE contribution creates a magnetoelectric range for which the surface wave propagation is forbidden, providing an additional signature of surface-polariton propagation in bi-isotropic media with an AHE contribution.

\end{abstract}

\maketitle
	
	\section{Introduction}

In general, the interaction of electromagnetic waves and collective excitations in matter can give rise to hybrid light-matter states or polaritons. When confined to an interface, they are described as surface polaritons (SPs), i.e., surface electromagnetic waves that propagate along the interface and decay away from it \cite{Cottam, Hopfield, Huang, Maradudin, Grosso}. Depending on the media composing the interface, different surface polaritons can emerge, including, for instance, plasmon-polaritons at metal or semiconductor surfaces \cite{Albuquerque}, surface phonon-polaritons in polar dielectrics \cite{Mills, Gubbin, Foteinopoulou}, magnon-polaritons  \cite{Bhoi,Bauer, Lehmeyer, Macedo, Hao}, and surface polaritons in nonlinear and quantum systems \cite{Saeid-Zorgabad1, Saeid-Zorgabad2, Saeid-Zorgabad4}.

Surface plasmon polaritons (SPPs) are infrared or visible-frequency excitations that arise from the coupling between an electromagnetic wave and collective electron oscillations traveling confined at the interface between two media (in general a dielectric and a metal), whose real dielectric constants have opposite signs \cite{refLUKAS,Agranovich,Polo-Mackay}. The excitation involves both electron motion in the metal surface (plasmon) and an electromagnetic wave in the dielectric substrate (polariton).
The SPP electric field is contained within the plane of propagation, while the magnetic field is perpendicular to this plane and parallel to the interface, thus constituting a transverse magnetic (TM) mode. SPPs at the interface of anisotropic media have also been a topic of great interest \cite{Haststein,Wallis,Camley},  as they exist for several parameter configurations not allowed at isotropic interface media \cite{Warmbier, Golenitskii}. In these systems, losses naturally arise due to damping factors (e.g., complex electric permittivity), which capture dissipative effects.

On the other hand, surface modes propagating without attenuation can occur in particular scenarios. For instance, Dyakonov surface waves are lossless excitations that propagate at the interface between two dielectric media, with at least one of them anisotropic \cite{Dyakonov}. Its simplest version occurs at the interface between an isotropic dielectric (refractive index $n_1$) and a uniaxial crystal (with refractive indices $n_e$ and $n_o$ fo the extraordinary and ordinary rays), when it holds $n_e>n_i>n_o$, in a limited angular existence domain (AED).  Richer Dyakonov waves arise when the patterning materials become more complex, as in the case of a uniaxial/uniaxial interface \cite{Polo}, isotropic/biaxial interface \cite{iso-biaxial}, and biaxial/biaxial interfaces \cite{bi-biaxial}. Additional patterning structures and properties of Dyakonov waves are found in reviews \cite{Review1,Review2}.

Surface waves are particularly sensitive to changes in the dielectric properties at material interfaces due to their strong field confinement near the boundary. This high sensitivity makes them powerful probes of interfacial phenomena, allowing their broad application in chemical and biological sensing technologies based on the surface plasmon resonance (SPR) technique \cite{Homola,Mayer,Kravets}.

SPPs have broad applications in technological devices in connection with the chemistry and physics of materials and surfaces. Identifying their occurrence in unusual materials is a topical issue for matter characterization. SPPs have been examined in new scenarios, such as the interface of topological insulators \cite{Qi,Chang, Karch}, three-dimensional topological insulators described by a dynamical axion field \cite{Zhu}, and topological Weyl semimetal contexts~\cite{Hofmann}.

Anisotropy effects also appear in matter ruled by bi-anisotropic constitutive relations,
	\begin{equation}
		D^{i}    =\epsilon_{ij}E^{j}+\alpha_{ij}B^{j }, \quad 	H^{i}    =\xi_{ij}B^{j}+\beta_{ij}E^{j}, \label{CRBAn}
	\end{equation}
with $\alpha_{ij}$ e $\beta_{ij}$ being magnetoelectric parameters which describe a broad class of materials with diverse optical properties. For permittivity, permeability, and magnetoelectric parameters written as diagonal isotropic matrices,  $\epsilon_{ij}=\epsilon\delta_{ij}$,	$\xi_{ij}=\mu^{-1}\delta_{ij}$, $\alpha_{ij}=\alpha\delta_{ij}$,  $\beta_{ij}=\beta\delta_{ij}$, 
the relations (\ref{CRBAn}) reduce to the bi-isotropic constitutive relations \cite{Sihvola,Kong},
\begin{align}
	\mathbf{D} = \epsilon \mathbf{E} + \alpha \mathbf{B}, \quad 
	\mathbf{H} = \dfrac{1}{\mu} \mathbf{B} + \beta \mathbf{E},
	\label{CRBi1}
\end{align}
where the complex coefficients satisfy $\beta^* = -\alpha$ to ensure energy conservation \cite{Sihvola1}. In bi-isotropic media, the electric and magnetic fields are linearly coupled in a way that does not give rise to privileged wave propagation directions. These materials have attracted much attention in connection with topological insulators \cite{Chang1,Lakhtakia}, axion electrodynamics \cite{Urrutia,Urrutia2}, characterization of materials \cite{Aladadi}, and birefringence effects \cite{PedroPRB02022}.

As far as we know, the conditions for the existence of SPPs in the interface of a chiral bi-isotropic medium have been investigated since 2005 \cite{Jin}, presenting distinguishing features in a chiral-metal interface, such as the presence of an s-wave and the dependence of the propagation length on the chiral parameter \cite{MiVan}. Richer scenarios involving chiral matter and anisotropic dielectric interfaces were also considered, with the attainment of surface Dyakonov-like waves \cite{Noonan}. The increasing of the propagation length with the chirality factor was confirmed in Ref. \cite{Seulong}, which examined a bi-isotropic layer in the Kretschmann configuration as well.  Chiral interfaces remain of great interest from both theoretical and applied perspectives \cite{Naheed,Pellegrini,LinPRL,KimKim}.

In this work, we investigate surface polaritons (SPs) at the interface between an isotropic dielectric and a bi-isotropic medium, both in the absence and presence of loss in the first medium. After briefly revisiting the basics on SPs propagation properties at a conventional dielectric/dielectric interface in Sec.~\ref{section-2}, we address the surface waves at the interface between an isotropic dielectric (medium 1) and a bi-isotropic dielectric (medium 2) in Sec.~\ref{SP3}. We discuss the enhanced possibilities for the existence of SPs in parameter space, as well as the propagation length and wavelength of the solutions in the presence of loss. In Sec.~\ref{SPAHE}, we consider an interface involving a bi-isotropic medium with an AHE term, both in the absence and presence of loss in medium 1. We identify multiple existence regimes and analyze how the AHE factor modifies the properties of the resulting bi-isotropic SP waves. In Sec.~\ref{conclusion}, we present our final remarks.

\section{\label{section-2}Basics on surface modes} 

The propagation of surface electromagnetic waves is described by applying suitable boundary conditions on the Maxwell equations in continuous matter. The simplest configuration occurs in a usual isotropic dielectric/dielectric interface disposed along the plane x-y, with boundary conditions
\begin{subequations}
	\begin{align}
		\mathbf{D}_{1}\cdot\mathbf{n}  &  =\mathbf{D}_{2}\cdot\mathbf{n}, \label{BCDn}\\
		\mathbf{E}_{1}\times\mathbf{n}  &  =\mathbf{E}_{2}\times\mathbf{n,} \label{BCEt}
	\end{align} 
\end{subequations}
the $z$-axis is orthogonal to the interface, $\mathbf{n}=\hat{z}$. Here, the labels $1$ and $2$ refer to medium 1 and 2, respectively.  The boundary conditions (\ref{BCEt}) and (\ref{BCDn}) yield
	\begin{subequations}
		\begin{align}
			E_{1x}-E_{2x}  &  =0, \label{Cond1a}\\
			\epsilon_{1}E_{1z}-\epsilon_{2}E_{2z}  &  =0, \label{Cond1b}
		\end{align}
	\end{subequations} 
	since $\mathbf{n}\cdot\mathbf{B}=0$. On the other hand, the Gauss's law,  $\mathbf{k}\cdot\mathbf{D}=0$, in both media, implies
	\begin{subequations}
		\begin{align}
			k_{x}E_{1x}+k_{1z}E_{1z}  &  =0, \label{Cond2a}\\
			k_{x}E_{2x}+k_{2z}E_{2z}  &  =0. \label{Cond2b}
		\end{align}
\end{subequations}
The p-polarized electrical field for the wave traveling along the interface is
\begin{equation}
	\mathbf{E}_{j}=\left(
	\begin{array}
		[c]{c}%
		E_{jx}\\
		0\\
		E_{jz}%
	\end{array}
	\right)  e^{i(k_{x}x+k_{jz}z-\omega t)}, \label{Efieldj}
\end{equation}
where $j=1,2$ labels the two media \cite{refLUKAS}. The wave vector, $	k^{2}=k_{x}^{2}+k_{jz}^{2}$, fulfills
	\begin{equation}
k_{x}^{2}+k_{jz}^{2}=\mu_j \epsilon_j \omega^2. \label{k2}
\end{equation}
The associated magnetic field, $\boldsymbol{H}=\frac{1}{\mu\omega}\mathbf{k}\times\mathbf{E}$, taking into account the electric field (\ref{Efieldj}), is 
	\begin{equation}\label{Hfield1}	
		\boldsymbol{H}_{j}=\frac{1}{\mu_j\omega}\left(
		\begin{array}
			[c]{c}%
			0\\
			k_{jz}E_{0}-k_{x}E_{jz}\\
			0
		\end{array}
		\right)  e^{i(k_{x}x+k_{jz}z-\omega t)},
	\end{equation}
being along the y-axis, that is, orthogonal to the interface (TM mode). With $\mu_1=\mu_2=1$, these equations provide 
\begin{equation}\label{kxusual1}
k_{x}^{2} =\left(  \frac{\epsilon_{1}\epsilon_{2}}{\epsilon_{1}+\epsilon_{2}}\right)  \mu_{0}\omega^{2},
\end{equation}
and the components along the z-axis,
\begin{align}\label{kzusual1}
	k_{1z}^{2}&=\left(  \frac{\epsilon_{1}^{2}}%
	{\epsilon_{1}+\epsilon_{2}}\right)  \mu_{0}\omega^{2}, \quad 
k_{2z}^{2} =\left(  \frac{\epsilon_{2}^{2}}{\epsilon
		_{1}+\epsilon_{2}}\right)  \mu_{0}\omega^{2},
\end{align}
which imply confined propagating waves at the boundary, $z=0$, when $k_{1z}$ and $k_{2z}$ are both purely imaginary and $k_{x}$ is real. It takes place for
\begin{itemize}
	\item[i)] $\epsilon_{1}<0$,  $\epsilon_{2}>0$, $|\epsilon_1|>\epsilon_{2}$, 
	\item[ii)] $\epsilon_{1}>0$ $\epsilon_{2}<0$ and $|\epsilon_2|>\epsilon_{1}$.
\end{itemize}
This simplified description of surface waves provides a starting point for examining SPs in more involved scenarios, as shown below.

\section{\label{SP3}SP solutions at bi-isotropic interfaces}

We now consider an interface between an isotropic dielectric (medium 1) and a bi-isotropic dielectric (medium 2), whose constitutive relations are
\begin{equation}
		\mathbf{D} = \epsilon_2\mathbf{E} + \alpha\mathbf{B}, \qquad
		\mathbf{H} = \dfrac{1}{\mu_2}\mathbf{B} + \beta\mathbf{E},
		 \label{constitutive-relations}
\end{equation}
with $\beta^* = -\alpha$. The electric permittivity tensor takes the form
\begin{equation}
	\bar{\epsilon}_{ij}    =\epsilon_{2}\delta_{ij}+\frac{\left(  \alpha
		+\beta\right)  }{\omega}\epsilon_{ijl}k_{l}, 
\end{equation}
Supposing that $\alpha$ and $\beta$ are complex,  $\alpha$, $\beta \in \mathbb{C}$, one has
\begin{equation}\label{MELF}
	\alpha=\alpha' +\mathrm{i} \alpha'', \quad \beta=\beta' +\mathrm{i} \beta'',
\end{equation}
where $\alpha'=\mathrm{Re}[\alpha]$, $\alpha''=\mathrm{Im}[\alpha]$, $\beta'=\mathrm{Re}[\beta]$ and $\beta''=\mathrm{Im}[\beta]$. The condition $\beta^* = -\alpha$ implies
\begin{equation}
	\alpha'=-\beta', \quad \alpha''=\beta'',
\end{equation}
so that  $\alpha+\beta= 2{\mathrm{i}\alpha''}$. The wave equation provides two wave vector solutions
\begin{equation}
	k_\pm^2=\left[  \mu_{2}\epsilon_{2}-2Z\right]  \omega
	^{2}\mp 2\mu_{2}\omega^{2}\alpha''  \sqrt{\mu_{2}%
		\epsilon_{2}-Z},
\end{equation}
where $Z=\mu_{2}^{2}\alpha''^2$. Such an equation yields two real and positive refractive indices,
\begin{equation} 
	n_{\pm }=\sqrt{\mu_2 \epsilon_2 +\mu ^{2}{\alpha ^{\prime \prime }}^{2}}\mp \mu_2
	\alpha ^{\prime \prime },  \label{isotropic-case-1-2}
\end{equation}%
which are associated with LCP and RCP waves, providing circular birefringence.

For such an interface, the boundary conditions are the same ones that hold for the usual case, given in Eqs.~(\ref{Cond1a}), (\ref{Cond1b}), (\ref{Cond2a}), (\ref{Cond2b}). This is related to the fact that the electric field and wave vector do not have components in the \( \hat{y} \) direction, whereas the magnetic field has only a y-component (TM mode). Thus, all contributions arising from the magnetoelectric parameters follow directly from the dispersion relation and refractive indices. As medium 2 presents two different refractive indices, there are two distinct solutions for $k^2$,
\begin{equation}
k_{x,\pm}^{2}+k_{2z,\pm}^{2}=\left[  \mu_{2}\epsilon_{2}-2Z\right]  \omega
^{2}\pm i\mu_{2}\omega^{2}\left(  \alpha+\beta\right)  \sqrt{\mu_{2}\epsilon
	_{2}-Z},
\end{equation}
that imply two distinct expressions for $k_x$ and $k_{jz}$, stemming from the system (\ref{Cond1a}), (\ref{Cond1b}), (\ref{Cond2a}), (\ref{Cond2b}).  
In medium 1, the dispersion relation is
\begin{equation}
k_{x,\pm}^{2}+k_{1z,\pm}^{2}=\mu_{1}\epsilon_{1}\omega^{2},
\end{equation}
where we have used Eq.~(\ref{k2}). As the components \( k_{x,\pm} \) must be the same in both media, we find
\begin{subequations} \label{k2BIinterface}
	\begin{equation}
		k_{x,\pm}^{2}    =\frac{\omega^{2}\epsilon_{1}\epsilon_{2}}{\epsilon_{2}%
			^{2}-\epsilon_{1}^{2}}\left\{ \epsilon_{2}-\epsilon_{1}%
		-\frac{2\epsilon_{1}\alpha^{\prime\prime2}}{\epsilon_{2}}\left[
		1\mp\sqrt{\frac{\epsilon_{2}}{\alpha^{\prime\prime2}}+1}\right]
		\right\},
	\end{equation}
	\begin{equation}
		k_{iz,\pm}^{2}    =\frac{\omega^{2}\epsilon_{i}^{2}}{\epsilon_{2}^{2}%
			-\epsilon_{1}^{2}}\left\{ \epsilon_{2}-\epsilon_{1}+2\alpha^{\prime\prime2}\left[  1\mp\sqrt{\frac{\epsilon_{2}}{\alpha^{\prime\prime2}}+1}\right]  \right\},
	\end{equation}
\end{subequations}
with $k_{1z}$ and $k_{2z}$ representing the wave vector in medium 1 (isotropic dielectric) and medium 2 (bi-isotropic dielectric), respectively. The double sign $\pm$ is ascribed to the two refractive indices (\ref{isotropic-case-1-2}).  Here, for simplicity, we have set $\mu_1=\mu_2 = 1$.

\subsection{\label{lossless-bi-isotropic-SP-section}Lossless bi-isotropic surface polaritons}

As is known, the existence of surface waves requires that $k_{i,z}$ be purely imaginary. In usual scenarios (simple dielectrics), such a condition is fulfilled when the permittivities have opposite signs, as shown in the requirements (i) and (ii) at the end of Sec.~\ref{section-2}.  However, on the surface of a bi-isotropic medium, this requirement is relaxed, and the surface polariton can emerge even for real and positive dielectric functions, $\epsilon_{1}, \epsilon_{2}$. In fact, imposing that $\epsilon_{1}, \epsilon_{2}$, and $\alpha^{\prime\prime}$ are real and positive, one finds
\begin{equation}
\epsilon_{2}<\epsilon_{1}<\epsilon_{2}+2\alpha^{\prime \prime 2}(1+\sqrt{1+\epsilon_{2}/\alpha^{\prime\prime 2}}), \label{condee1a}
\end{equation}
for the $k_{-}$ mode, and
\begin{equation}
	\epsilon_{2}+2\alpha^{\prime \prime2}(1-\sqrt{1+\epsilon_{2}/\alpha^{\prime\prime 2}})<\epsilon_{1}<\epsilon_{2},  \label{condee2a}
\end{equation}
for the $k_{+}$ mode. 

\begin{figure}[h]
	\centering
	\includegraphics[scale=.5]{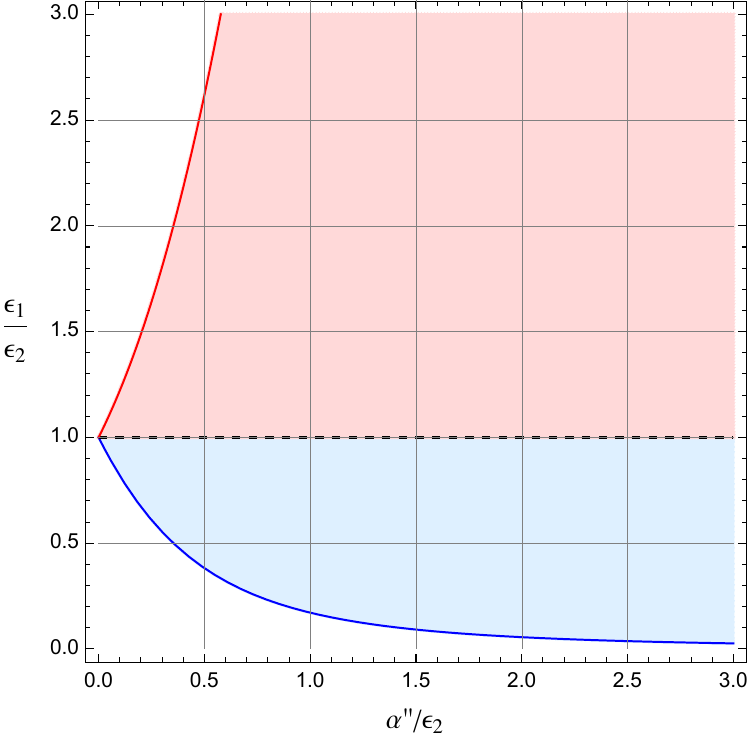}
	\caption{\justifying\small{\label{plot-lossless-bi-isotropic-polariton} Shaded region for the existence of SPs in bi-isotropic media in the space $\epsilon_1/\epsilon_2 \times \alpha^{\prime\prime}/\epsilon_2$. Here, the red (blue) sections depict the minus (plus) solutions related to Eqs.~(\ref{condee1a}) and (\ref{condee2a}), respectively. }}
\end{figure}

Equations (\ref{condee1a}) and (\ref{condee2a}) assure the realization of surface waves for several parameter values. Such amplification of possibilities is initially illustrated in Fig.~\ref{plot-lossless-bi-isotropic-polariton}, which displays the allowed region (in the electromagnetic parameter space) for the existence of lossless SPs for both positive permittivities.

The two shaded parameter regions (red and blue) involve positive permittivities, $\epsilon_{1}>0, \epsilon_{2}>0$,  and an additional condition on the bi-isotropic parameter, as pointed out in the first two lines of Tabs.~\ref{tab:conditions_SPP_lossless_mode_plus} and \ref{tab:conditions_SPP_lossless_mode_minus}. For positive permittivities, surface waves occur only for one mode in each region in the parameter space. For $\epsilon_{1} <\epsilon_{2}$ (blue shaded region), surface waves emerge just for the plus mode, since only it can provide nonnull $\alpha^{\prime\prime}$.  On the other hand, for $\epsilon_{1} >\epsilon_{2}$ (red shaded region), only the minus mode propagates. See the first two rows of Tabs.~\ref{tab:conditions_SPP_lossless_mode_plus} and \ref{tab:conditions_SPP_lossless_mode_minus}. Thus, each specific set of parameters lies in one of the regions (red or blue) of Fig.~\ref{plot-lossless-bi-isotropic-polariton}, which defines two distinct domains for SP emergence.
There are also forbidden regions where the surface modes do not occur (see the white region on the left of the red curve and below the blue line). Moreover, it is important to note that the new SP solutions predicted in Fig.~\ref{plot-lossless-bi-isotropic-polariton}, despite not requiring negative permittivities, can not be classified as Dyakonov waves, since they do propagate in limited angular window and do not have hybrid polarization.  

Besides the existence conditions for lossless non-Dyakonov SPs with positive/real electric permittivities $\epsilon_{1}$ and $\epsilon_{2}$, there are several other sets of parameters, including negative permittivity values that assure the appearance of SPs with a nonnull magnetoelectric coefficient.  For instance, when $\epsilon_{1} <\epsilon_{2}$ and  $\epsilon_2 < 0$, it holds
	\begin{align}
		\alpha^{\prime\prime 2}  \geq |\epsilon_{2}| , \label{condee1b}
	\end{align}
which may be useful for determining the values of the bi-isotropic parameter for such a configuration.

There also arises the peculiar SP scenario in which both permittivities are negative, $\epsilon_{1}<0$ and $\epsilon_{2}<0$, which occurs for $\alpha^{\prime\prime} \geq \sqrt{|\epsilon_{2}|}$. It suggests the possibility of surface modes propagation in metal/metal-like bi-isotropic platforms, a system that deserves further investigation in the future.
See Tabs.~\ref{tab:conditions_SPP_lossless_mode_plus} and \ref{tab:conditions_SPP_lossless_mode_minus} for the general conditions on the plus and minus SP modes, respectively.

For the cases involving negative permittivities, the occurrence of both plus and minus modes is guaranteed provided that $\alpha^{\prime\prime}\neq 0$ and the corresponding conditions given in rows 3, 4, and 5 of Tables \ref{tab:conditions_SPP_lossless_mode_plus} and \ref{tab:conditions_SPP_lossless_mode_minus} hold for each particular case.

\begin{table}[H]
	\centering
	\caption{{Conditions on $\alpha^{\prime\prime}$ for SPs (mode $+$).}}
	\label{tab:conditions_SPP_lossless_mode_plus}
	\begin{tabular}{*{4}{c}}
		\toprule[0.8pt]\midrule
		Permittivity restrictions & & &  Bi-isotropic term condition   \\
		\midrule
	 $\epsilon_{1}>0$, $\epsilon_{2}>0$, $\epsilon_{1} < \epsilon_{2}$  &  && $ \displaystyle \alpha^{\prime\prime} > \frac{1}{2} \frac {|\epsilon_{2}-\epsilon_{1}| }{\sqrt{\epsilon_{1}}}$   \\
 $\epsilon_{1}>0$, $\epsilon_{2}>0$, 	 $\epsilon_{1} \geq \epsilon_{2}$ & & & --- \\
 $\epsilon_{1}>0$, $\epsilon_{2}<0$,  & & & $\epsilon_{1} < - \epsilon_{2}$ and $ \displaystyle \alpha^{\prime\prime} > \frac{1}{2} \frac {|\epsilon_{2}-\epsilon_{1}| }{\sqrt{\epsilon_{1}}}$ \\
  $\epsilon_{1}<0$, $\epsilon_{2}>0$ & & & $\epsilon_{1} < - \epsilon_{2}$ and $\alpha^{\prime \prime}>0$ \\
  $\epsilon_{1}<0$, $\epsilon_{2}<0$ & & & $\epsilon_{1} < \epsilon_{2}$ and $\alpha^{\prime \prime} \geq \sqrt{ - \epsilon_{2}}$  \\
	 		\midrule\bottomrule[0.8pt]	
	\end{tabular}
\end{table}	

\begin{table}[H]
	\centering
	\caption{{Conditions on $\alpha^{\prime\prime}$ for SPs (mode $-$).}}
	\label{tab:conditions_SPP_lossless_mode_minus}
	\begin{tabular}{*{4}{c}}
		\toprule[0.8pt]\midrule
		Permittivity restrictions & & &  Bi-isotropic term condition   \\
		\midrule
	 $\epsilon_{1}>0$, $\epsilon_{2}>0$, $\epsilon_{1} \leq \epsilon_{2}$  &  && ---  \\
 $\epsilon_{1}>0$, $\epsilon_{2}>0$, 	 $\epsilon_{1} > \epsilon_{2}$ & & & $\displaystyle{ \alpha^{\prime\prime} > \frac{1}{2} \frac{|\epsilon_{1}- \epsilon_{2}|}{\sqrt{\epsilon_{1}}} }$  \\
 $\epsilon_{1}>0$, $\epsilon_{2}<0$,  & & & $\epsilon_{1} > - \epsilon_{2}$ and $ \displaystyle \alpha^{\prime\prime} > \frac{1}{2} \frac {|\epsilon_{2}-\epsilon_{1}| }{\sqrt{\epsilon_{1}}}$ \\
  $\epsilon_{1}<0$, $\epsilon_{2}>0$ & & & $\epsilon_{1} < - \epsilon_{2}$ and $\alpha^{\prime \prime}>0$ \\
  $\epsilon_{1}<0$, $\epsilon_{2}<0$ & & & $\epsilon_{1} < \epsilon_{2}$ and $\alpha^{\prime \prime} \geq \sqrt{ - \epsilon_{2}}$  \\
	 		\midrule\bottomrule[0.8pt]	
	\end{tabular}
\end{table}

Finally, we emphasize that the conditions (\ref{condee1a}) and (\ref{condee2a}) and all the cases in Tabs.~\ref{tab:conditions_SPP_lossless_mode_plus} and \ref{tab:conditions_SPP_lossless_mode_minus} hold for nonnull magnetoelectric coefficient, $\alpha^{\prime\prime}=0$. Thus, a bi-isotropic medium offers a platform supporting lossless surface polaritons without the requirement of negative permittivities or anisotropic permittivities (as in the case of Dyakonov surface waves \cite{Dyakonov,Polo, iso-biaxial,bi-biaxial}).

\subsection{\label{SPP-bi-isotropic-with-losses-section}SPP's in metal/bi-isotropic interfaces with attenuation}

Up to this point, we have considered lossless cases. Now, we turn our attention to investigating scenarios of lossy propagation \cite{refLUKAS}. For that, we consider an interface between a lossless bi-isotropic medium 2 and a lossy medium 1, like a metal, whose permittivity is
\begin{equation}
	\epsilon_{1}=\epsilon_{1}'+\mathrm{i}\epsilon_{1}''
\end{equation}
with $\epsilon_{1}'$ and $\epsilon_{1}''$ being real parameters. Keeping $\epsilon_2$ real and working in the scenario of low loss, $\epsilon_{1}''\ll \epsilon_{1}'$, the wave vector component $k_x$ acquires both real and imaginary parts, $k=k'+\mathrm{i}k''$, being given as

\begin{subequations} 
\label{kMBiinterface1}
\begin{equation}
	k_{x,\pm}=k_0	\sqrt{
		\frac{\epsilon_{1}^{\prime\,2} m_\pm-\epsilon_{1}^{\prime}\epsilon_{2}^{2}}{\epsilon_{1}^{\prime\,2}-\epsilon_{2}^{2}} }\left[1+\mathrm{i} \frac{\epsilon_{2}^{2}\epsilon_{1}^{\prime\prime}N_{1\pm}}{\eta
	\left(	\epsilon_{2}^{2}-\epsilon_{1}^{\prime}m_\pm	\right)	} \right] 
	\label{kx}
\end{equation}
\begin{align}
k_{1z,\pm}&=k_0	\sqrt{
		\frac{	\epsilon_1^{\prime 2} \left(\epsilon_1' - m_\pm\right) }{	\epsilon_1'^2 - \epsilon_2^2	} }\left[1+\mathrm{i}	\frac{ \epsilon''_{1} N_{2\pm}}{\eta \left(\epsilon_1'-m_\pm\right)} \right] , \label{k1z} \\
	k_{2z,\pm}&=k_0\sqrt{
		\frac{
			\epsilon_{2}^{2}\left(\epsilon_{1}^{\prime}-m_\pm\right)
		}{\epsilon_{1}^{\prime 2}-\epsilon_{2}^{2}
		}
	}\left[1+\mathrm{i}	\frac{\epsilon_{1}^{\prime} \epsilon''_{1} N_{1\pm}}{\eta
	\left(\epsilon_{1}^{\prime}-m_\pm\right)} \right] ,  	\label{k2z}
\end{align}
\end{subequations}
with
\begin{subequations}
\begin{align}
	m_{\pm}&= \epsilon_2\mp 2\sqrt{\alpha''^{\,2}\left(\alpha''^{\,2}+\epsilon_2\right)} + 2\alpha''^{\,2}, \\
	N_{1\pm}&= (\epsilon_{2}^{2}-\epsilon_{1}^{\prime}m_{\pm}+\epsilon_{1}^{\prime 2} ) , \\
	 N_{2\pm}&=-2\epsilon_2^{2}m_\pm - \epsilon_1'^3 +3\epsilon_1'\epsilon_2^2, \\
	\eta &= 	2\epsilon_{1}^{\prime}\left( \epsilon_{2}^{2}-\epsilon_{1}^{\prime 2} \right), 
\end{align}
\end{subequations}
where the signs ($\pm$) are associated with $k_{\pm}$ solutions and \(k_0 = {2\pi}/{\lambda_0}\), with \(\lambda_0\) being the wavelength associated with the adopted value of \(\epsilon_{1}\). 
Considering a simple dielectric ($\alpha'' = 0$), we have $m_\pm = \epsilon_{2}$, and, after some algebra, the result for an interface standard dielectric/ lossy dielectric is recovered.

For the case of small losses, the condition for the existence of surface plasmon polaritons is given by
\begin{align}
\epsilon_{2}+ 2\alpha^{\prime \prime2}(1-\sqrt{1+\epsilon_{2}/\alpha^{\prime \prime 2}})< \epsilon_{1}^{\prime} < \epsilon_{2} , \label{condition-SPP-with-losses-mode-plus-1}
\end{align}
for the plus mode, and
\begin{align}
\epsilon_{2} < \epsilon_{1}^{\prime} < \epsilon_{2} + 2 \alpha^{\prime\prime 2} (1+\sqrt{1+\epsilon_{2}/\alpha^{\prime\prime 2}}) , \label{condition-SPP-with-losses-mode-minus-1}
\end{align}
for the minus mode, which defines the allowed regions in the parameter space for positive values of the parameters.

The SPP wavelength ($\lambda_{\text{SPP}}$) and propagation length ($L_{\text{SPP}}$) can be determined from the real ($k'_x$) and imaginary ($k''_x$) parts of the wave vector \cite{refLUKAS}, respectively, and are given by 
\begin{equation}
	L_{\mathrm{SPP}}=\dfrac{1}{2k''_x}, \qquad  \lambda_{\mathrm{SPP}}=\dfrac{2\pi}{k'_x}. \label{definition-lengths-polaritions-1}
\end{equation}

\begin{figure}[h]
	\centering
\includegraphics[scale=.6]{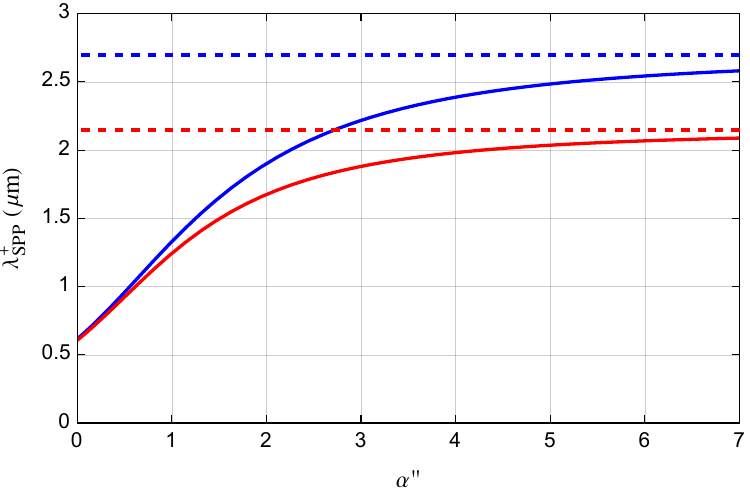}
	\caption{ \justifying\small{Behavior of $\lambda_{SPP}^{+}$ as a function of the magnetoelectric parameter $\alpha''$ for silver, $\epsilon_1 = -18.2 + 0.5i$ (blue curve) and for gold, $\epsilon_1 = -11.6 + 1.2i$ (red curve). Here, we have used $\epsilon_{2}=1$. The horizontal dashed lines are given by \eqref{assymptotic-wavelength-plus}. } }
	\label{SPP-lambda}
\end{figure}

\begin{figure}[h]
	\centering
\includegraphics[scale=.6]{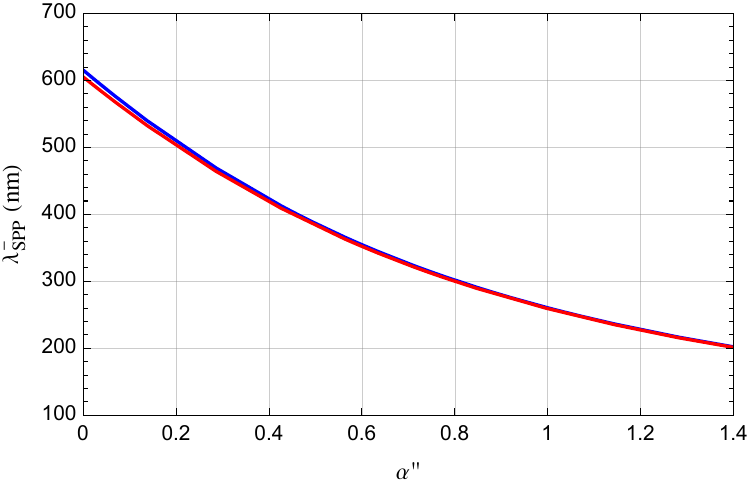}
	\caption{ \justifying\small{Behavior of $\lambda_{SPP}^{-}$ as a function of the magnetoelectric parameter $\alpha''$ for silver (blue curve) and for gold (red curve). Here, we have used $\epsilon_{2}=1$.} }
	\label{SPP-lambda-modo-menos-modelo-2}
\end{figure}

In Figs.~\ref{SPP-lambda} and \ref{SPP-lambda-modo-menos-modelo-2}, we illustrate the general behavior of $\lambda_{SPP}$ associated with each mode described in \eqref{kx} in terms of the magnetoelectric parameter $\alpha^{\prime\prime}$. The blue curve corresponds to silver ($\epsilon_1 = -18.2 + 0.5i$) and the red curve to gold ($\epsilon_1 = -11.6 + 1.2i$), for $\lambda = 633$ nm \cite{refLUKAS}.  For the plus mode, see Fig.~\ref{SPP-lambda}, starting from the $\alpha'' = 0$ value (ordinary dielectric), the wavelength $\lambda_{\mathrm{SPP}}^{+}$ increases with $\alpha^{\prime\prime}$ until reaching the assymptotic value
\begin{align}
\lambda^{+}_{\mathrm{M}} &= \frac{2\pi}{\omega} \sqrt{ \frac{ \epsilon_{2}^{2}-\epsilon_{1}^{\prime 2}}{\epsilon_{1}^{\prime} \epsilon_{2}^{2}}}.  \label{assymptotic-wavelength-plus}
\end{align}

On the other hand, in Fig.~\ref{SPP-lambda-modo-menos-modelo-2}, we observe no significant difference between plasmon-polariton wavelengths for the minus mode (for silver and gold) as the bi-isotropic term $\alpha^{\prime\prime}$ increases. Additionally, in the limit of very large $\alpha^{\prime\prime}$, it holds $\lambda_{\mathrm{SPP}}^{-} \rightarrow 0$, indicating an increasingly large wavevector for this mode.

\begin{figure}[h]
	\centering
	\centering\includegraphics[scale=.6]{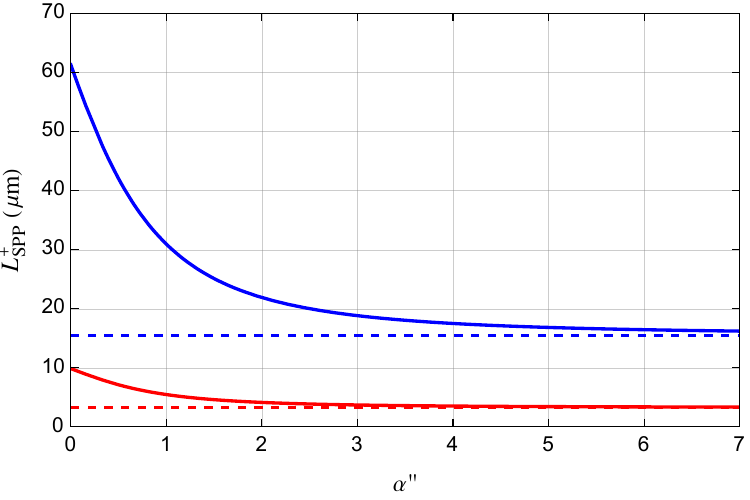}
	\caption{\justifying\small{Propagation length $L_{\mathrm{SPP}}^{+}$ in terms of $\alpha''$ for silver (blue curve) and for gold (red curve). Here, we have used $\epsilon_{2}=1$. The horizontal dashed curves are given by \eqref{assymptotic-wavelength-mode-plus}.}}
	\label{SSP-length}
\end{figure}

\begin{figure}[H]
	\centering
	\centering\includegraphics[scale=.6]{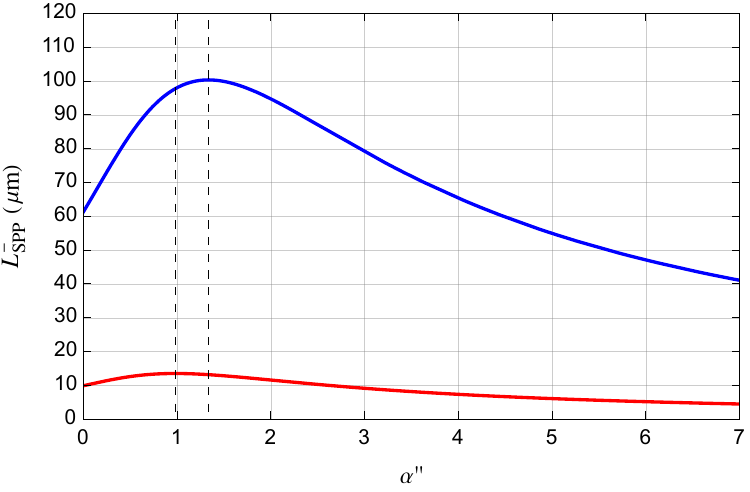}
	\caption{\justifying\small{Propagation length $L_{\mathrm{SPP}}^{-}$ in terms of $\alpha''$ for silver (blue curve) and for gold (red curve). Here, we have used $\epsilon_{2}=1$. The vertical dashed lines are given by $\hat{\alpha}$ defined in \eqref{alpha-critic-for-propagation-length-mode-minus}.}}
	\label{SSP-length-modo-minus}
\end{figure}

Considering now the propagation length of \eqref{definition-lengths-polaritions-1}, its general behavior is depicted in Figs.~\ref{SSP-length} and \ref{SSP-length-modo-minus}, for $L_{\mathrm{SPP}}^{\pm}$, respectively, in terms of the magnetoelectric parameter $\alpha^{\prime\prime}$. For the plus mode, the propagation length decreases monotonically with $\alpha^{\prime\prime}$, starting from the usual value at $\alpha^{\prime\prime} = 0$. In the theoretical limit  of a very large magnetoelectric parameter, one finds
\begin{align}
L^{+}_{\mathrm{m}} &= \frac{ (\epsilon^{\prime 2}_{1}- \epsilon_{2}^{2})^{2}}{\omega \epsilon^{\prime\prime}_{1} \epsilon_{2}^{2} (\epsilon^{\prime 2}_{1} + \epsilon_{2}^{2})} \sqrt{ \frac{ \epsilon_{2}^{2} \epsilon^{\prime}_{1}}{\epsilon_{2}^{2} - \epsilon_{1}^{\prime 2}}} , \label{assymptotic-wavelength-mode-plus}
\end{align}
which indicates that the larger $\epsilon_{1}^{\prime\prime}$, the smaller the propagation length, as indicated by the colored dashed lines in Fig.~\ref{SSP-length}, for silver (blue) with $\epsilon^{\prime\prime}_{1} = 0.5$ and gold (red) with $\epsilon^{\prime\prime}_{1}=1.2$.

For the minus mode, illustrated in Fig.~\ref{SSP-length-modo-minus}, one notes that, differently from the plus mode, the propagation length reaches a peak at a specific value of the magnetoelectric parameter, 
\begin{align}
\hat{\alpha} &= \frac{( \epsilon_{1}^{\prime} - \epsilon_{2}) \, (\epsilon_{1}^{\prime}+ 3\epsilon_{2})}{2 \sqrt{ 6 \epsilon_{1}^{\prime} \epsilon_{2}^{2} - 2 \epsilon_{1}^{\prime 3}}}, \label{alpha-critic-for-propagation-length-mode-minus}
\end{align}
defined for
\begin{align}
\epsilon_{2} >0, \, \, \mathrm{and} \, \, \epsilon_{1}^{\prime} < - 3 \epsilon_{2} . \label{condition-critical-alpha-1}
\end{align}
Such an unusual maximum, occurring at $\hat{\alpha}$, as far as we know, has still not been reported in the literature and can be regarded as a signature of the surface plasmon-polaritons characterized by the minus mode ($k_{x,-}$). For $\epsilon_{1}^{\prime} > - 3 \epsilon_{2}$, $L_{SPP}^{-}$ behaves similarly to the propagation length of Fig.~\ref{SSP-length}. It is important to remark that this feature takes place when $\epsilon_{1}^{\prime} < - 3 \epsilon_{2}$ for positive $\epsilon_{2}$. On the other hand, for $\epsilon_{2}<0$, the the peak at $\hat{\alpha}^{\prime\prime}$ happens when the condition $\epsilon_{1}^{\prime} < 3 \epsilon_{2}$ holds. Moreover, in the regime of very large $\alpha$, it holds $L^{-}_{\mathrm{SPP}} \rightarrow 0$, indicating that the SPP becomes strongly attenuated with increasing values of the bi-isotropic parameter.

According to Figs.~\ref{SPP-lambda} and \ref{SSP-length}, we observe that the lossy material with larger $\epsilon^{\prime}_{1}$ (silver) is the one whose wavelength undergoes major amplification and propagation length displays a larger reduction (as $\alpha^{\prime\prime}$ increases).


Finally, it is worth mentioning that the peak in the propagation length can also arise in the plus mode. Indeed, for $\epsilon_{2}>0$, there occurs a peak in $L_{SPP}^{+}$ at
\begin{align}
\alpha^{\prime\prime}=\frac{(\epsilon_{2}-\epsilon_{1}^{\prime})  |\epsilon_{1}^{\prime}+3\epsilon_{2}|}{2 \sqrt{6\epsilon_{1}^{\prime} \epsilon_{2}^{2} - 2 \epsilon_{1}^{\prime 3}}}, \label{critical-alpha-SPP-bi-isotropic-with-losses-mode-plus-1}
\end{align}
 for $ -3 \epsilon_{2} < \epsilon_{1}^{\prime} < - \sqrt{3} \epsilon_{2}$.

\section{\label{SPAHE}Interfacial waves in bi-isotropic AHE-dielectric interfaces}

We now consider an interface between a conventional lossless dielectric (medium 1) and a bi-isotropic medium with an anomalous Hall current (medium 2), recently investigated in the literature \cite{Alex}. For the latter, the modified Maxwell equations,
\label{maxwell-equations-plane-wave-1}
\begin{align}
	\mathbf{k}\cdot\mathbf{D}   =  -i\mathbf{b} \cdot \mathbf{B} , \quad \quad \mathbf{k}\times\mathbf{H}   =-\omega\mathbf{D} + i	\,	\mathbf{b}\times\mathbf{E},   \label{maxwell-equations-1} 
\end{align}
with the same constitutive relations (\ref{constitutive-relations}),  yield the refractive indices
\begin{align}
	\label{RCPLCP} 
	{n_{2,\pm}} &=\mu\alpha^{\prime\prime}\pm \sqrt{ \mu \epsilon +\mu^2\alpha^{\prime\prime2} - \mu b/\omega}, \\
{\tilde{n}_{2,\pm} } &  =-\mu\alpha^{\prime\prime}\pm \sqrt{ \mu \epsilon +\mu^2\alpha^{\prime\prime2} + \mu b/\omega},
\end{align}
for the configuration in which $\mathbf{n}$ is parallel to the vector $\mathbf{b}$. See details in Ref.~\cite{Alex}. Henceforth, only the refractive indices ${n_{2,\pm}}$ will be used in the following analysis.

\subsection{Free attenuation case}
It is worth mentioning that the aforementioned boundary conditions (\ref{Cond1a}), (\ref{Cond1b}), (\ref{Cond2a}), (\ref{Cond2b}) are not affected by the inclusion of the AHE term because we are considering the chiral vector along the propagation direction, $\mathbf{b} \parallel \mathbf{k}$, in such a way the Gauss’s law in the bi-isotropic dielectric with AHE term remains unchanged. Using these boundary conditions, the components of the wave vector $\mathbf{k}$ (for $n_{1,\pm}$) can be written as:
\begin{equation}\label{k2zCBISO}
	k_{iz,\pm}^{2}=\frac{\epsilon_{i}^{2}k_0^{2}}{\epsilon_{1}^{2}-\epsilon_{2}%
		^{2}}\left[  \epsilon_{1}-\alpha^{\prime\prime2}\left(  1\mp\sqrt
	{1+\frac{\epsilon_{2}\omega-b}{\alpha^{\prime\prime2}\omega}}\right)
	^{2}\right],
\end{equation}
and
\begin{equation}\label{k2xCBISO}
k_{x,\pm}^{2}=\frac{\epsilon_{2}\epsilon_{1}k_0^{2}}{\epsilon_{2}%
	^{2}-\epsilon_{1}^{2}}\left[  \epsilon_{2}-\frac{\epsilon_{1}}{\epsilon_{2}%
}\alpha^{\prime\prime2}\left(  1\mp\sqrt{1+\frac{\epsilon_{2}\omega-b}%
	{\alpha^{\prime\prime2}\omega}}\right)  ^{2}\right].
\end{equation}
As expected, it is simple to verify that for $b = 0$ we recover the expressions (\ref{k2BIinterface}), previously discussed.  As it occurs in (\ref{k2BIinterface}) and (\ref{kMBiinterface1}), the simultaneous presence of the magnetoelectric parameter, $\alpha^{\prime\prime}$, and the AHE factor, $b$, in relations (\ref{k2zCBISO}) and (\ref{k2xCBISO}), creates new possibilities for the appearance of SPs. Now, however, the scenario is still more involved, given the dependence on the frequency, which makes the existence conditions hold for specific frequency bands. Each row of these tables displays a relation analogous to the conditions (\ref{condee1a}) and (\ref{condee2a}) for the lossless bi-isotropic SPs.

\begin{table}[H]
	\caption{SPP conditions for the mode $(+)$ case in different frequency regimes.}
	\centering
	\label{Conditions_SPP_lossless_b_mode_k+}
	\setlength{\tabcolsep}{5.5pt}  \label{table-AHE-1}
	\begin{tabular}{*{3}{c}}
	\toprule\toprule
               &  \textbf{SPP conditions} & \textbf{Frequency band} \\
		\midrule
	\multirow{2}{*}{ $ \epsilon_{2} < \alpha^{\prime \prime 2}$} & $\epsilon_2 > \epsilon_1 > (\alpha''-Q_-)^2$
		&
		$ \omega > \omega_k \; $
		\\
		& $\epsilon_2 < \epsilon_1 < (\alpha''-Q_-)^2$
		& 
		$\omega_0 < \omega < \omega_k$ 
		\\
		\midrule
		\multirow{2}{*}{ $ \epsilon_{2} \geq \alpha^{\prime \prime 2}$} & $\epsilon_{2} > \epsilon_{1} > (\alpha^{\prime\prime} - Q_{-})^{2}$ &  $\omega_{0} < \omega < \omega_{R}$; $\omega > \omega_{R}$ \\
		& $0 < \epsilon_{1} < \epsilon_{2}$ & $\omega = \omega_{R}$ \\
		\bottomrule
	\end{tabular}
\end{table}

\begin{table}[H]
	\caption{SPP conditions for the mode $(-)$ case in different frequency regimes.}
	\centering
	\label{Conditions_SPP_lossless_b_mode_k-}
	\setlength{\tabcolsep}{8pt} \label{table-AHE-2}
	\begin{tabular}{*{3}{c}}
	\toprule\toprule
		& \textbf{SPP conditions} & \textbf{Frequency band} \\
		\midrule
	\multirow{1}{*}{ $ \epsilon_{2} \leq \alpha^{\prime \prime 2}$}  &	$\epsilon_2 < \epsilon_1 < (\alpha''+Q_-)^2 $ & $ \omega > \omega_{0} $ 
		\\
		\midrule
		\multirow{2}{*}{ $ \epsilon_{2} > \alpha^{\prime \prime 2}$} & $\epsilon_2 < \epsilon_1 < (\alpha''+Q_-)^2 $ &  $\omega > \omega_k $
		\\
		& $\epsilon_{2} > \epsilon_{1} > (\alpha^{\prime\prime} + Q_{-})^{2}$  & $\omega_{0} < \omega < \omega_{k}$ \\
		\bottomrule
	\end{tabular}
\end{table}

In these tables, there appear characteristic frequencies which are listed as follows:
\begin{equation}	\omega_0=\dfrac{b}{\epsilon_{2}+\alpha^{\prime\prime 2}}, \quad	\omega_R=\dfrac{b}{\epsilon_{2}}, \quad	\omega_k=\dfrac{b}{2\alpha^{\prime\prime}\sqrt{\epsilon_{2}}},
\end{equation}
and the factor,
\begin{equation}
Q_{-}=\sqrt{ \mu \epsilon +\mu^2\alpha^{\prime\prime2} - \mu b/\omega}.
\end{equation}

\begin{figure}
	\centering
	\begin{subfigure}{0.48\textwidth}
		\centering
		\includegraphics[width=\linewidth]{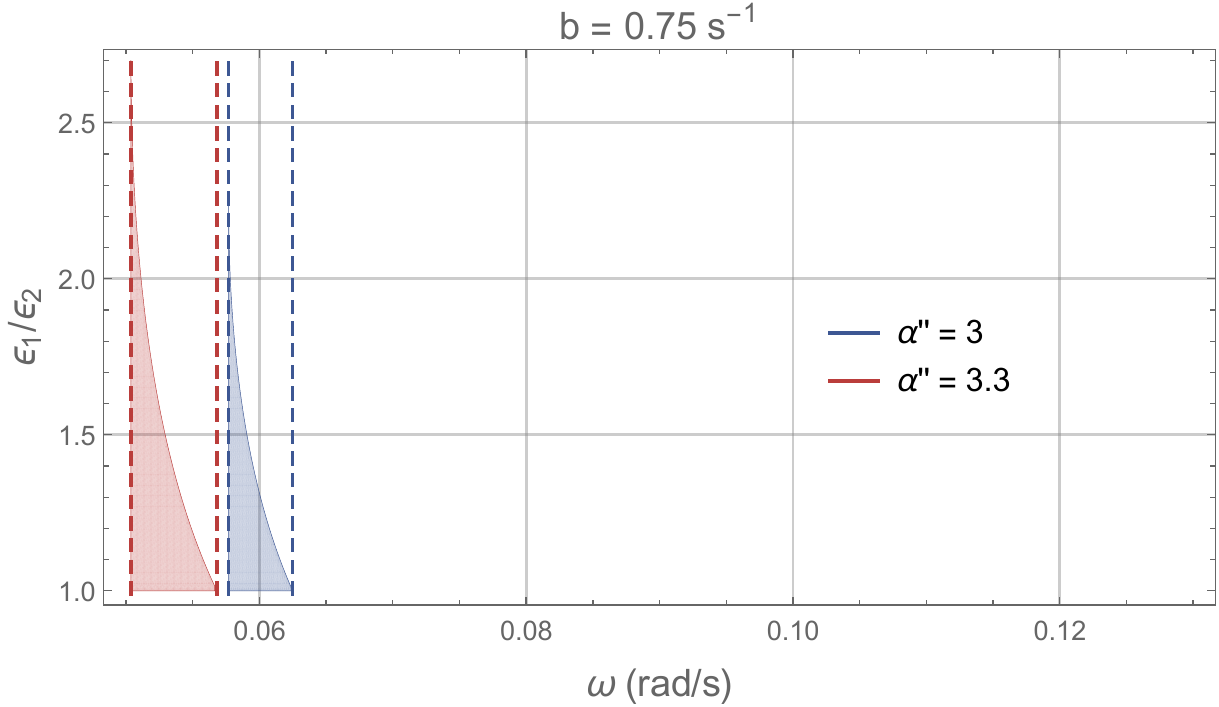}
		
		\label{AHE-Region1}
	\end{subfigure}
	\hfill
	\begin{subfigure}{0.48\textwidth}
		\centering
		\includegraphics[width=\linewidth]{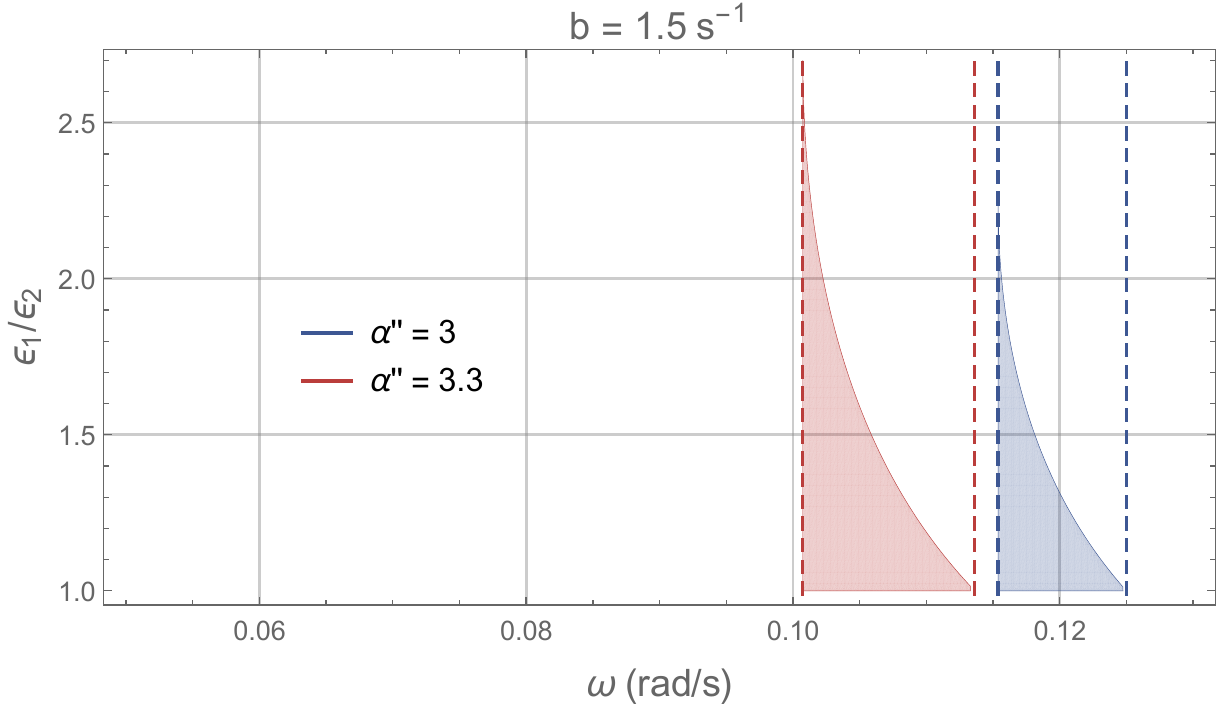}
		
		\label{AHE-Region2}
	\end{subfigure}
	
	\caption{\justifying SP existence region for the mode (+) at the region $\omega_0 < \omega < \omega_{k}$. Here, we observe how the magnetoelectric parameter $\alpha''$ and the AHE term $b$ affects the SP existence conditions. In this mode, increasing the magnetoelectric parameter broadens the SP existence region while shifting it toward lower frequencies, whereas the AHE term not only enlarges this region but also shifts it toward higher frequencies, enabling propagation at higher frequencies. Here,we have used $\epsilon_{2}=4$ and $\mu=1$}
	\label{fig:AHE-regions}
\end{figure}

\begin{figure}
	\centering
	\begin{subfigure}{0.48\textwidth}
		\centering
		\includegraphics[width=\linewidth]{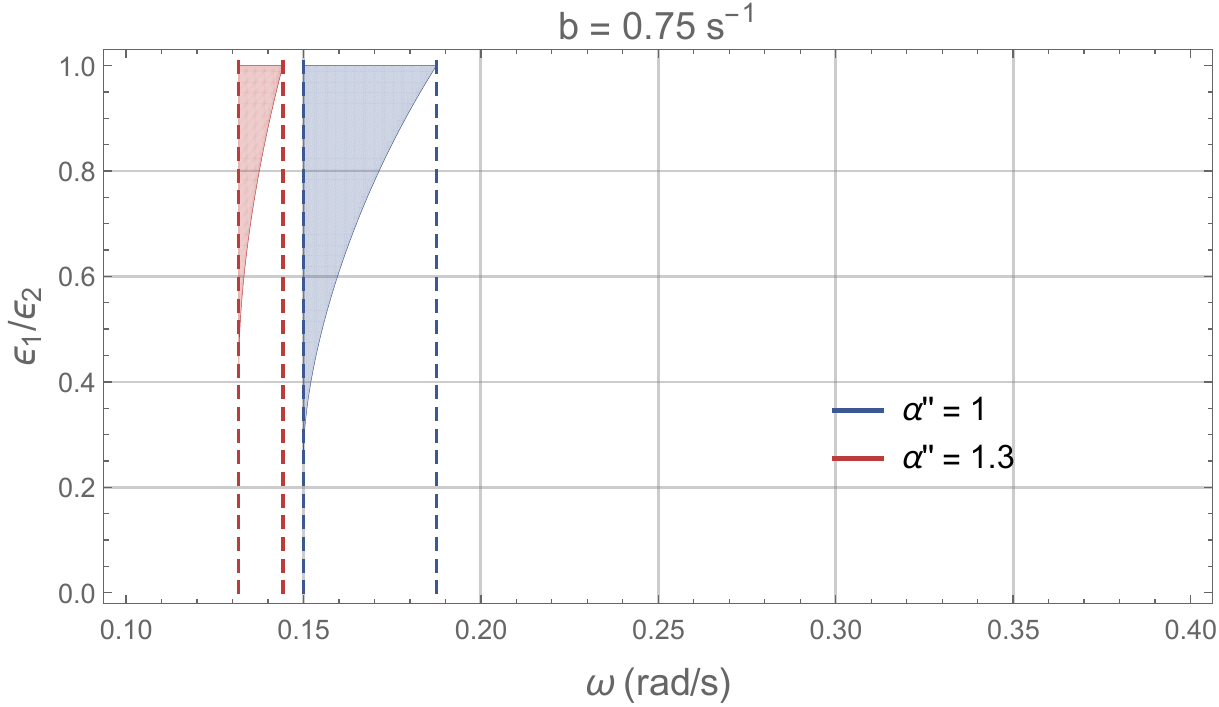}
		
		\label{AHE-Region3}
	\end{subfigure}
	\hfill
	\begin{subfigure}{0.48\textwidth}
		\centering
		\includegraphics[width=\linewidth]{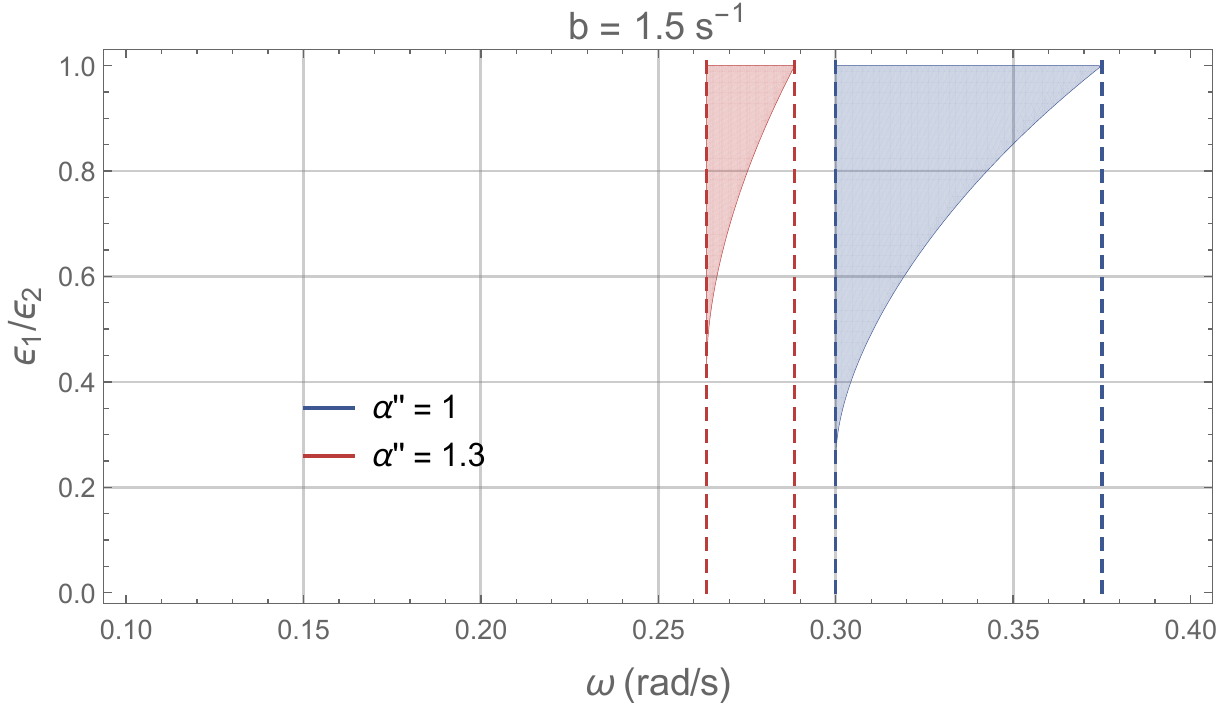}
		
		\label{AHE-Region4}
	\end{subfigure}
	
	\caption{\justifying SP existence region for the mode (-) at the region $\omega_0 < \omega < \omega_{k}$. Increasing the magnetoelectric parameter reduces the SP existence region and shifts it toward lower frequencies, whereas the effect of the AHE term remains unchanged.
		Here,we have used $\epsilon_{2}=4$ and $\mu=1$.}
	\label{fig:AHE-regions3}
\end{figure}
Considering the refractive index $n_{2,-}$, it is worth noting that it exhibits the peculiar feature of being negative for $\omega > \omega_R$ \cite{Alex}.  Although negative refraction includes nontrivial effects compared to a simple dielectric, it does not prevent the formation of confined modes at the interface, fully assuring the conditions for the occurrence of SPs in this frequency range.

From \figref{fig:AHE-regions} and \figref{fig:AHE-regions3}, we observe that the modes supported by the dielectric–bi-isotropic-medium interface, in the presence of an anomalous Hall current,  exhibit distinct behaviors, originated from the structure of \eqref{RCPLCP}. In \figref{fig:AHE-regions}, the existence regions are defined for $\epsilon_1>\epsilon_2$, whereas in \figref{fig:AHE-regions3} they occur for $\epsilon_1<\epsilon_2$, for the range 	$\omega_0 < \omega < \omega_k$, in accordance with  Tab.~\ref{table-AHE-1} (first row, second condition) and Tab.~\ref{table-AHE-2} (second row, second condition). Therefore, differently from the non-dispersive bi-isotropic surface polaritons of Sec.~\ref{lossless-bi-isotropic-SP-section}, the associated plus and minus modes can now exist at the same frequency range, in the two positive parameter regions $\epsilon_{1}>\epsilon_{2}$ and $\epsilon_{1}<\epsilon_{2}$, being one in each region. 

Another feature is the possibility of the mode minus and mode plus coexisting in the same parameter region. As an example, in the region defined by $\epsilon_1>\epsilon_2$, there is minus SP mode in the range $\omega_0 < \omega < \omega_k$ (see first row, second condition of Tab.~\ref{table-AHE-1}) and also a plus SP mode in the range $\omega> \omega_0 $ (see first row, first condition of Tab.~\ref{table-AHE-2}). In this situation, the frequency range is not the same. In the region defined by $\epsilon_2>\epsilon_1$, there is minus SP mode in the range $\omega_0 < \omega < \omega_R$ (see second row, first condition of Tab.~\ref{table-AHE-1}) and also a plus SP mode in the range $\omega_0 < \omega < \omega_k$  (see second row, second condition of Tab.~\ref{table-AHE-2}). 

It is also interesting to emphasize that for both modes, the magnetoelectric parameter $\alpha^{\prime\prime}$ and the AHE term $b$ shift the existence regions in opposite directions along the frequency axis. This two-parameter contrasting behavior yields rich and flexible existence conditions which constitute a characteristic signature of SPs at this particular type of interface.

In addition to the existence conditions for lossless non-Dyakonov surface polaritons with positive/real electric permittivities $\epsilon_{1}$ and $\epsilon_{2}$ in the presence of a positive bi-isotropic parameter and AHE term, discussed previously, it is also possible to obtain several conditions on the magnitude of $\alpha^{\prime\prime}$ ensuring the existence of SPs, provided that the permittivities and AHE term satisfy specific requirements. The different sets of conditions, including distinct combinations among the several parameters, are summarized in Tables \ref{tab:conditions_SPP_lossless_mode_plus_with_AHE} and \ref{tab:conditions_SPP_lossless_mode_minus_with_AHE} for the plus and minus modes, respectively.

For both modes, besides the possibility of coexistence in the two positive parameter regions $\epsilon_{1}>\epsilon_{2}$ and $\epsilon_{1}<\epsilon_{2}$ already mentioned, one can also find well-defined parameter regions and frequency ranges where the two modes propagate. Examining the first two rows in Tabs. \ref{tab:conditions_SPP_lossless_mode_plus_with_AHE} and \ref{tab:conditions_SPP_lossless_mode_minus_with_AHE}, one reports coexistence of both modes in three regions, namely

\begin{itemize}
\item[i)] for $\epsilon_{1} <\epsilon_{2}$, with
\begin{align}
0< \alpha^{\prime\prime} <\frac{1}{2\sqrt{\epsilon_{1}}} | (b/\omega) + (\epsilon_{1}-\epsilon_{2})|,
\end{align}
in the frequency range
\begin{align}
\frac{b}{\epsilon_{2}} \leq \omega < \frac{b}{\epsilon_{2} - \epsilon_{1}}, 
\end{align}
\item[ii)] for $\epsilon_{1} <\epsilon_{2}$, with
\begin{align}
\sqrt{-\epsilon_{2} + \frac{b}{\omega}} \leq \alpha^{\prime\prime} <\frac{1}{2\sqrt{\epsilon_{1}}} | (b/\omega) + (\epsilon_{1}-\epsilon_{2})|, 
\end{align}
in the frequency range
\begin{align}
\frac{b}{\epsilon_{1}+\epsilon_{2}} < \omega < \frac{b}{\epsilon_{2}} ,
\end{align}
\item[iii)] for $\epsilon_{1} >\epsilon_{2}$, with
\begin{align}
\sqrt{-\epsilon_{2} + \frac{b}{\omega}} \leq \alpha^{\prime\prime} <\frac{1}{2\sqrt{\epsilon_{1}}} | (b/\omega) + (\epsilon_{1}-\epsilon_{2})|,
\end{align}
in the frequency range
\begin{align}
0<\omega < \frac{b}{\epsilon_{1}+\epsilon_{2}}.
\end{align}
\end{itemize}

Such a behavior is absent in the corresponding case without the AHE term and can thus be regarded as a characteristic signature of the interplay between the magnetoelectric and anomalous Hall responses in the surface-polariton propagation. Nevertheless, for $b=0$, when one of the permittivities ($\epsilon_{1}$ or $\epsilon_{2}$) is negative (which may happen in metalic substrates for specific wavelengths), both modes are allowed to occur as described by rows 3, 4, and 5 of Tables \ref{tab:conditions_SPP_lossless_mode_plus} and \ref{tab:conditions_SPP_lossless_mode_minus}, considering that the corresponding conditions are satisfied.

\begin{widetext}	

	\begin{table}[H]
		\begin{center}
				\caption{{Conditions on $\alpha^{\prime\prime}$ for SPs (mode $+$) with nonnull AHE term.}}
				\label{tab:conditions_SPP_lossless_mode_plus_with_AHE}
				\setlength{\tabcolsep}{14pt}
				\begin{tabular}{*{3}{c}}
					\toprule[0.8pt]\midrule
					Permittivity restrictions &  Bi-isotropic term condition & Frequency range  \\
					\midrule
					\multirow{4}{*}{ $\epsilon_{1}>0$, $\epsilon_{2}>0$, $\epsilon_{1} < \epsilon_{2}$ }  &  $ \displaystyle \alpha^{\prime\prime} > \frac{1}{2 \sqrt{\epsilon_{1}}} | (b/\omega) + (\epsilon_{1}-\epsilon_{2})| $  & $0< \omega \leq \displaystyle{\frac{b}{\epsilon_{1}+\epsilon_{2}} }$ ; $\omega >  \displaystyle{\frac{b}{\epsilon_{2}-\epsilon_{1}}} $   \\[1.5ex]
					& $\alpha^{\prime\prime} >0$ & $ \displaystyle{\frac{b}{\epsilon_{2}} } \leq \omega \leq \displaystyle{ \frac{b}{\epsilon_{2}-\epsilon_{1}}}$ \\[1.5ex]
					& $\alpha^{\prime\prime} \geq \sqrt{-\epsilon_{2}+(b/\omega) }$ & $\displaystyle{ \frac{b}{\epsilon_{1}+\epsilon_{2}}} < \omega < \displaystyle{\frac{b}{\epsilon_{2}}}$ \\[1.5ex]
					\midrule 
					$\epsilon_{1}>0$, $\epsilon_{2}>0$, $\epsilon_{1} > \epsilon_{2}$	& $ \sqrt{-\epsilon_{2} + (b/\omega)} \leq \alpha^{\prime\prime} < \displaystyle{\frac{1}{2\sqrt{\epsilon_{1}}} } | (b/\omega) + (\epsilon_{1}- \epsilon_{2})| $ & $\omega < \displaystyle{\frac{b}{\epsilon_{1}+\epsilon_{2}}}$ \\[1ex]
					\midrule
					\multirow{3}{*}{ $\epsilon_{1}<0$, $\epsilon_{2}>0$, $\epsilon_{2} < -\epsilon_{1}$ }  & $\alpha^{\prime\prime} \geq \sqrt{-\epsilon_{2} + (b/\omega)}$ &  $0< \omega < \displaystyle{\frac{b}{\epsilon_{2}}}$ \\[1ex]
					& $\alpha^{\prime\prime} >0$ & $\omega \geq \displaystyle{ \frac{b}{\epsilon_{2}}}$ \\[1ex]
					\midrule
					$\epsilon_{1}>0$, $\epsilon_{2}<0$, $ \epsilon_{2} < - \epsilon_{1}$  & $\alpha^{\prime\prime} > \displaystyle{\frac{1}{2\sqrt{\epsilon_{1}}}} | (b/\omega) + (\epsilon_{1}-\epsilon_{2})|$  & $\omega >0$ \\[1ex]
					\midrule
					$\epsilon_{1}>0$, $\epsilon_{2}<0$, $\epsilon_{2} > - \epsilon_{1}$ & $\sqrt{-\epsilon_{2}+(b/\omega)} \leq \alpha^{\prime\prime} < \displaystyle{\frac{1}{2\sqrt{\epsilon_{1}}}} | (b/\omega) + (\epsilon_{1}-\epsilon_{2})|$ & $ 0 < \omega < \displaystyle{\frac{b}{\epsilon_{1}+\epsilon_{2}}} $ \\[1ex]
					\midrule
					$\epsilon_{1}<0$, $\epsilon_{2}<0$, $\epsilon_{1}<\epsilon_{2}$ & $\alpha^{\prime\prime} \geq \sqrt{-\epsilon_{2} + (b/\omega) } $ &  $ \omega > 0$ \\[1ex]
					\midrule\bottomrule[0.8pt]	
				\end{tabular}
		\end{center}
	\end{table}

	\begin{table}[h]
		\begin{center}
				\caption{Conditions on $\alpha^{\prime\prime}$ for SPs (mode $-$) with nonnull AHE term.}
				\label{tab:conditions_SPP_lossless_mode_minus_with_AHE}
				\setlength{\tabcolsep}{22pt}
				\begin{tabular}{*{3}{c}}
					\toprule[0.8pt]\midrule
					Permittivity restrictions &  Bi-isotropic term condition & Frequency range  \\
					\midrule
					\multirow{3}{*}{ $\epsilon_{1}>0$, $\epsilon_{2}>0$, $\epsilon_{1} < \epsilon_{2}$ }  &  $ \sqrt{  - \epsilon_{2} + (b/\omega) } \leq \alpha^{\prime \prime} < \displaystyle {\frac{1}{2\sqrt{\epsilon_{1}}}}  | (b/\omega) + (\epsilon_{1}-\epsilon_{2}) | $  &  $ \displaystyle{ \frac{b}{\epsilon_{1}+\epsilon_{2}} } < \omega < \displaystyle{ \frac{b}{\epsilon_{2}}} $   \\[1.5ex]
					& $\alpha^{\prime\prime} < \displaystyle{ \frac{1}{2\sqrt{\epsilon_{1}}}} | (b/\omega) + (\epsilon_{1}-\epsilon_{2}) | $ & $\displaystyle{ \frac{b}{\epsilon_{2}}} \leq \omega < \displaystyle{ \frac{b}{\epsilon_{2}-\epsilon_{1}}}$ \\[1.5ex]
					\midrule
					\multirow{4}{*}{ $\epsilon_{1}>0$, $\epsilon_{2}>0$, $\epsilon_{1} > \epsilon_{2}$ }  & $\alpha^{\prime\prime} \geq \sqrt{-\epsilon_{2} + (b/\omega)} $ & $0<\omega < \displaystyle{ \frac{b}{\epsilon_{1}+\epsilon_{2}}}$ \\	[1.5ex]	
					& $\alpha^{\prime\prime} > \sqrt{-\epsilon_{2} + (b/\omega)}$ & $\omega = \displaystyle{ \frac{b}{\epsilon_{1}+\epsilon_{2}}}$ \\[1.5ex]
					& $\alpha^{\prime\prime} > \displaystyle{\frac{1}{2\sqrt{\epsilon_{1}}}} | (b/\omega) + (\epsilon_{1}- \epsilon_{2})|$ & $\omega > \displaystyle{\frac{b}{\epsilon_{1}+\epsilon_{2}}}$ \\[1ex]
					\midrule 
					\multirow{3}{*}{ $\epsilon_{1}<0$, $\epsilon_{2}>0$, $\epsilon_{2} < - \epsilon_{1}$ }  &  $\alpha^{\prime\prime} \geq \sqrt{-\epsilon_{2} + (b/\omega)}$ & $0 < \omega < \displaystyle{ {b}/\epsilon_{2}}$ \\[1ex]
					& $\alpha^{\prime\prime} >0$ & $\omega \geq \displaystyle{{b}/{\epsilon_{2}}}$ \\[1ex]
					\midrule
					\multirow{4}{*}{ $\epsilon_{1}>0$, $\epsilon_{2}<0$, $\epsilon_{2} > - \epsilon_{1}$ }  &  $\alpha^{\prime\prime} \geq \sqrt{-\epsilon_{2} + (b/\omega)}$ & $0 < \omega < \displaystyle{ \frac{b}{\epsilon_{1}+\epsilon_{2}}}$ \\[1.5ex]
					& $\alpha^{\prime\prime} > \sqrt{-\epsilon_{2} + (b/\omega)}$ & $\omega = \displaystyle{\frac{b}{\epsilon_{1}+\epsilon_{2}}}$ \\[1.5ex]
					& $\alpha^{\prime\prime} > \displaystyle{\frac{1}{2\sqrt{\epsilon_{1}}}} | (b/\omega) + (\epsilon_{1}- \epsilon_{2})|$ & $\omega > \displaystyle{\frac{b}{\epsilon_{1}+\epsilon_{2}}}$ \\[1.ex]
					\midrule
					$\epsilon_{1}<0$, $\epsilon_{2}<0$, $\epsilon_{1}<\epsilon_{2}$ & $\alpha^{\prime\prime} \geq \sqrt{-\epsilon_{2} + (b/\omega)}$ & $\omega>0$ \\[1.3ex]
					\bottomrule[0.8pt]	
				\end{tabular}
		\end{center}
	\end{table}	
\end{widetext}

\subsection{Attenuation case}

For the case where medium 1 is a metal and medium 2 is the bi-isotropic medium with an anomalous Hall current,  the momentum $k_{x,\pm}$ of Eq.~(\ref{k2xCBISO}) is rewritten as
	\begin{equation}
		\tilde{k}_{x,\pm}=k_0	\sqrt{
			\frac{\epsilon_{1}^{\prime\,2} M_\pm-\epsilon_{1}^{\prime}\epsilon_{2}^{2}}{\epsilon_{1}^{\prime\,2}-\epsilon_{2}^{2}} }\left[1+\mathrm{i} \frac{\epsilon_{2}^{2}\epsilon_{1}^{\prime\prime}\tilde{N}_{1\pm}}{\eta
			\left(	\epsilon_{2}^{2}-\epsilon_{1}^{\prime}M_\pm	\right)	} \right], 
		\label{kx-AHE}
	\end{equation}
where
\begin{equation}
	M_{\pm}=-b/\omega+
	\epsilon_2
	\mp2\sqrt{\alpha''^{\,2}\left(-b/\omega+\alpha''^{\,2}+\epsilon_2\right)}
	+ 2\alpha''^{\,2},
\end{equation}
and
\begin{equation}
	\tilde{N}_{1\pm}=  \epsilon_{2}^{2}-\epsilon_{1}^{\prime}M_{\pm}+\epsilon_{1}^{\prime 2} .
\end{equation}

Here we can write the propagation length, $L_{\mathrm{SPP}}$, and wavelength,  $\lambda_{\mathrm{SPP}}$, in the same way as in \eqref{definition-lengths-polaritions-1}, that is,
\begin{equation}
	L_{\mathrm{SPP}}=\dfrac{1}{2\tilde{k}''_x},  \quad  \quad \lambda_{\mathrm{SPP}}=\dfrac{2\pi}{\tilde{k}'_x}. \label{wavelength-equation-1}
\end{equation}

\begin{figure}[H]
	\centering
	\includegraphics[scale=.60]{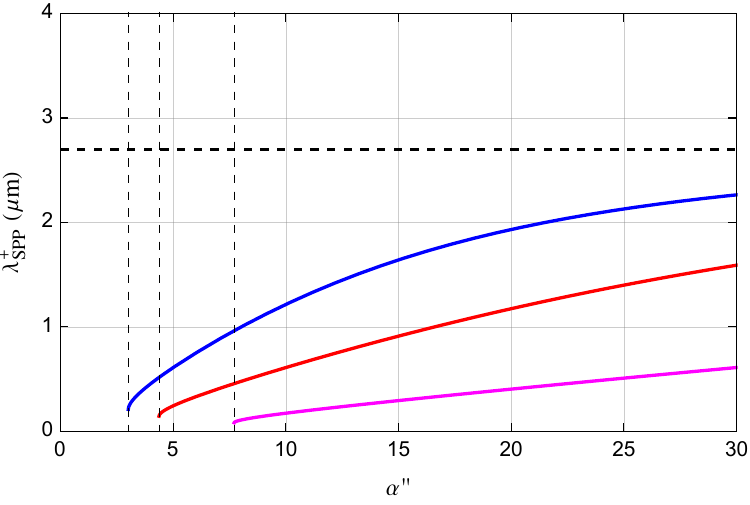}
	\caption{\justifying Wavelength $\lambda_{SPP}^{+}$ of \eqref{wavelength-equation-1} in terms of magnetoelectric parameter $\alpha^{\prime\prime}$. Here, we have  silver for $\epsilon_{1}$, $\epsilon_{2}=1$, and $b= 0.1$ (blue), $b=0.2$ (red), and $b=0.6$ (magenta). The vertical dashed lines are given by \eqref{alpha-bar-definition-1} for each example.}
	\label{plot-wavelength-AHE-mode-plus}
\end{figure}

\begin{figure}[H]
	\centering
	\includegraphics[scale=.60]{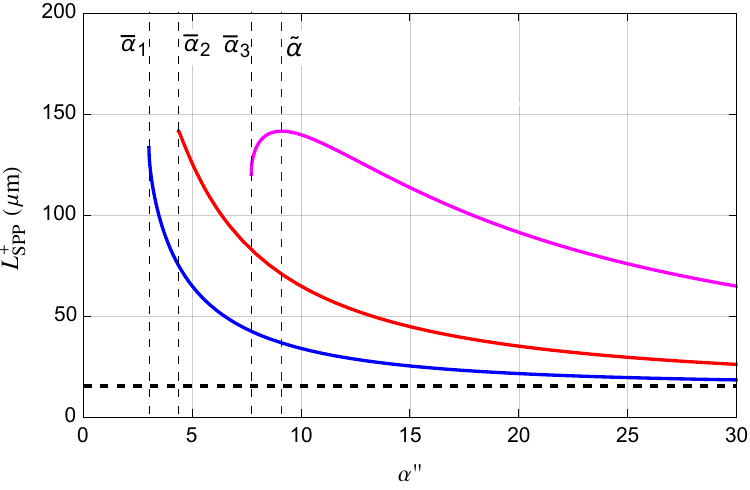}
	\caption{\justifying Propagation length $L_{SPP}^{+}$ of \eqref{wavelength-equation-1} in terms of magnetoelectric parameter $\alpha^{\prime\prime}$. Here, we have silver for $\epsilon_{1}$, and $\epsilon_{2}=1$, with $b= 0.1$ (blue), $b=0.2$ (red), and $b=0.6$ (magenta). The vertical dashed lines are given $\bar{\alpha}_{i}$ of \eqref{alpha-bar-definition-1}, with $i=\{1$ (blue), 2 (red), 3 (magenta)$\}$ representing each curve. The peak of the magenta curve occurs at $\tilde{\alpha}$ of \eqref{alpha-tilde-definition-1}.}	
	\label{plot-propagation-length-AHE-mode-plus}
\end{figure}

The results of Sec.~\ref{SP3} can be obtained from $M_{\pm}$ and $\tilde{N}_{1\pm}$ by setting $b=0$. Considering that medium 1 is silver (with $\epsilon_{1}=-18.2+0.2 i$, for $\lambda=633$ nm), the general behavior of the characteristic lengths is depicted in Figs.~\ref{plot-wavelength-AHE-mode-plus} and \ref{plot-propagation-length-AHE-mode-plus} for the plus mode.

In Fig.~\ref{plot-wavelength-AHE-mode-plus}, one observes that the SPP wavelength increases monotonically with $\alpha^{\prime\prime}$. In the regime of a very large bi-isotropic term, it goes to its asymptotic value $L_{m}^{+}$, given in \eqref{assymptotic-wavelength-mode-plus}, represented by the horizontal dashed line. Interestingly, the presence of non-null AHE term $b$ shifts the values of bi-isotropic parameter $\alpha^{\prime\prime}$ that yield the characteristic lengths (see the vertical dashed lines), now defined by $\alpha^{\prime\prime} > \bar{\alpha}$, with $\bar{\alpha}$ given by
\begin{align}
\bar{\alpha}&= \sqrt{(b/\omega)-\epsilon_{2}}. \label{alpha-bar-definition-1}
\end{align}
Such behavior does not happen in the scenario with $b=0$, as previously illustrated in Sec.~\ref{SPP-bi-isotropic-with-losses-section}.

In Fig.~\ref{plot-propagation-length-AHE-mode-plus}, we note that the propagation length $L_{SPP}^{+}$ decreases with $\alpha^{\prime\prime}$, in general. However, it is also possible that $L_{SPP}^{+}$ reaches a maximum value at $\alpha^{\prime\prime}=\tilde{\alpha}$, with $\tilde{\alpha}$ defined by
\begin{align}
\tilde{\alpha}&= \frac{ |( b/\omega) \epsilon_{1}^{\prime} - (\epsilon_{1}^{\prime 2} + \epsilon_{1}^{\prime} \epsilon_{2} - \epsilon_{2}^{2})  |}{2\sqrt{\epsilon_{1}^{\prime} (\epsilon_{2}^{2} - \epsilon_{1}^{\prime 2})}} , \label{alpha-tilde-definition-1}
\end{align}
which holds for $\epsilon_{1}<0$, $\epsilon_{2}>0$, and $\epsilon_{1}<-\epsilon_{2}$, and 
\begin{align}
0 < \omega < \frac{ b \epsilon_{1}^{\prime}}{\epsilon_{2}^{2} + \epsilon_{1}^{\prime} \epsilon_{2} - \epsilon_{1}^{2}}.  \label{frequency-range-peak-L-plus-AHE}
\end{align}
The forbidden regions that appear in the behavior of the wavelength, Fig.~\ref{plot-wavelength-AHE-mode-plus}, are also present in the characteristic length plot, exhibiting enlarged width with the magnetoelectric parameter. The existence of these regions can be regarded as a signature of polaritonic propagation in systems composed of bi-isotropic media with an AHE contribution.

\begin{figure}[H]
	\centering
	\includegraphics[scale=.60]{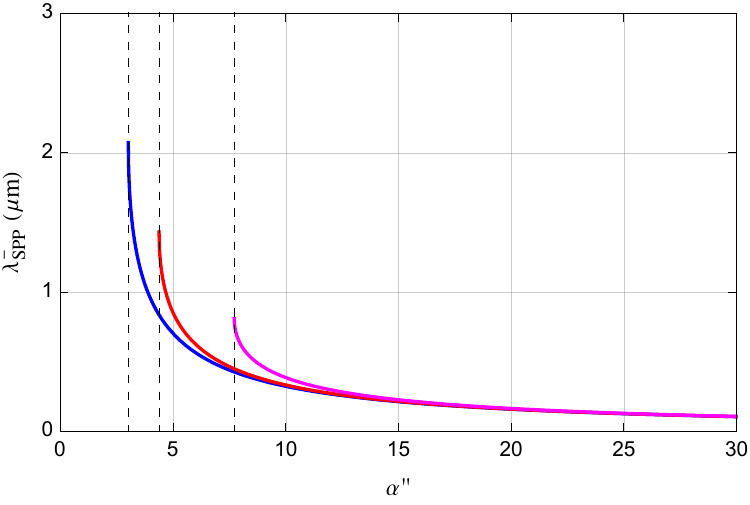}
	\caption{\justifying Wavelength $\lambda_{SPP}^{-}$ of \eqref{wavelength-equation-1} in terms of magnetoelectric parameter $\alpha^{\prime\prime}$. Here, we have  silver for $\epsilon_{1}$, $\epsilon_{2}=1$, and $b=0.1$ (blue), $b=0.2$ (red), $b=0.6$ (magenta). The vertical dashed lines are given by \eqref{alpha-bar-definition-1} for each example.} 
	\label{plot-wavelength-AHE-mode-minus}
\end{figure}

\begin{figure}[H]
	\centering
	\includegraphics[scale=.60]{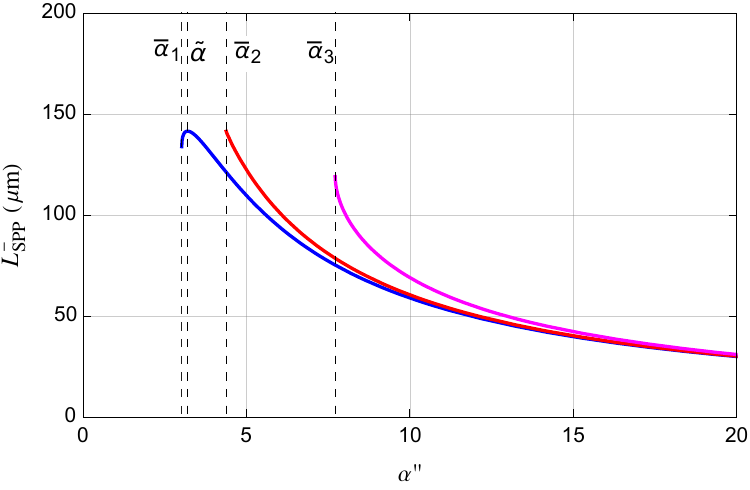}
	\caption{\justifying Propagation length $L_{SPP}^{-}$ of \eqref{wavelength-equation-1} in terms of magnetoelectric parameter $\alpha^{\prime\prime}$. Here, we have silver for $\epsilon_{1}$, $\epsilon_{2}=1$, and $b= 0.1$ (blue), $0.2$ (red), and $0.6$ (magenta). The vertical dashed lines are given $\bar{\alpha}_{i}$ of \eqref{alpha-bar-definition-1}, with $i=\{1$ (blue), 2 (red), 3 (magenta)$\}$ representing each curve. The peak of the blue curve is at $\tilde{\alpha}$ of \eqref{alpha-tilde-definition-1}.}	
	\label{plot-propagation-length-AHE-mode-minus}
\end{figure}

Considering the minus mode, the characteristic lengths \eqref{wavelength-equation-1} are depicted in Figs.~\ref{plot-wavelength-AHE-mode-minus} and 
\ref{plot-propagation-length-AHE-mode-minus} in terms of $\alpha^{\prime\prime}$. The wavelength decreases as $\alpha^{\prime\prime}$ increases, approaching its asymptotic value of zero for very large values of the bi-isotropic parameter. Regarding the propagation length $L_{SPP}^{-}$ of Fig.~\ref{plot-propagation-length-AHE-mode-minus}, one observes that this characteristic length also diminishes with $\alpha^{\prime\prime}$ for each example illustrated. However, the blue curve shows a peak in $L_{SPP}^{-}$. In fact, this peak occur at $\alpha^{\prime\prime} = \tilde{\alpha}$, with $\tilde{\alpha}$ given in \eqref{alpha-tilde-definition-1} for $\epsilon_{1}^{\prime}<0$, $\epsilon_{2}>0$, $\epsilon_{1}^{\prime} < - \epsilon_{2}$, and  
	\begin{align}
		\frac{b \epsilon_{1}^{\prime}}{\epsilon_{2}^{2}+ \epsilon_{1}^{\prime}\epsilon_{2}-\epsilon_{1}^{\prime 2}} < \omega \leq \frac{b}{\epsilon_{2}}. \label{frequency-range-peak-L-minus-AHE}
\end{align}

Therefore, the occurrence of a maximum in the propagation length also arises for the minus mode in a different frequency range. This behavior indicates that the interplay between the bi-isotropic parameter and the AHE contribution can enhance the propagation length within specific parameters, constituting an additional characteristic signature of the associated chiral surface-polaritons.

\section{Final remarks \label{conclusion}}

In this work, we have investigated the conditions for the emergence of TM surface polaritons (SPs) in chiral scenarios, including conventional bi-isotropic dielectrics and bi-isotropic matter with anomalous Hall current (AHE).

In Sec.~\ref{lossless-bi-isotropic-SP-section}, we have discussed the conditions for the existence of lossless surface polaritons at the interface between a conventional dielectric and a bi-isotropic medium. The magnetoelectric parameter $\alpha^{\prime\prime}$ relaxes the conventional requirement of a negative permittivity, allowing lossless SPs to exist without a contiguous metallic medium (for positive permittivities). As a consequence, the parameter space supporting SPs is substantially enlarged, giving rise to multiple branches where they do not coexist simultaneously. In particular, the interface supports lossless non-Dyakonov SPs for both real/positive $\epsilon_{1}>0, \epsilon_{2}>0$. Several configurations of SPs for real negative permittivities are also reported, as summarized in Tabs.~\ref{tab:conditions_SPP_lossless_mode_plus} and \ref{tab:conditions_SPP_lossless_mode_minus}.

Section \ref{SPP-bi-isotropic-with-losses-section} addresses surface polaritons at the interface between a conventional lossy medium (metal) and a bi-isotropic medium. The corresponding dispersion relations are derived, with existence conditions presented in \eqref{condition-SPP-with-losses-mode-plus-1} and \eqref{condition-SPP-with-losses-mode-minus-1}. In this case, the SP propagation length ($L_{\mathrm{SPP}}$) and the wavelength ($\lambda_{\mathrm{SPP}}$) are illustrated for some examples, revealing that the propagation length diminishes, while the SPP wavelength grows with the magnetoelectric parameter. Furthermore, it was noticed that the propagation length $L_{SPP}^{-}$ can first increase before decaying, exhibiting a peak at the value $\hat{\alpha}$, see \eqref{alpha-critic-for-propagation-length-mode-minus}, as illustrated in Fig.~\ref{SSP-length-modo-minus}. This peculiar behavior (ocurrence of a peak in the propagation length) also holds for the SP plus mode at the specific value $\alpha^{\prime\prime}$, see \eqref{critical-alpha-SPP-bi-isotropic-with-losses-mode-plus-1}, under the condition $ -3 \epsilon_{2} < \epsilon_{1}^{\prime} < - \sqrt{3} \epsilon_{2}$. Such an unusual propagation length enhancement can be taken as a signature of the bi-isotropic SPs.

Section \ref{SPAHE} extends the analysis by considering an interface between a conventional lossless dielectric (medium 1) and a bi-isotropic medium with an anomalous Hall current (medium 2). The corresponding dispersion relations are given in Eqs.~(\ref{k2zCBISO}) and (\ref{k2xCBISO}), while general conditions of existence for SPs are presented in Tables \ref{Conditions_SPP_lossless_b_mode_k+} and \ref{Conditions_SPP_lossless_b_mode_k-}. The interplay between the magnetoelectric parameter and the AHE term considerably extends the existence conditions, multiplying the parameter regions where SPs can propagate, within characteristic cutoff frequencies, as presented in 
Tables \ref{tab:conditions_SPP_lossless_mode_plus_with_AHE} and \ref{tab:conditions_SPP_lossless_mode_minus_with_AHE}. In fact, new existence possibilities are induced by the AHE factor in the bi-isotropic scenario: i) The bi-isotropic plus and minus non-Dyakonov surface polaritons modes can now exist at the same frequency range, in the two positive parameter regions $\epsilon_{1}>\epsilon_{2}$ and $\epsilon_{1}<\epsilon_{2}$, being one in each region; ii)  the mode minus and the mode plus can coexist in the same parameter region, but not the same frequency range; iii) for both modes, besides the possibility of coexistence in the two positive parameter regions $\epsilon_{1}>\epsilon_{2}$ and $\epsilon_{1}<\epsilon_{2}$, there are well defined parameter regions and frequency ranges where the two modes propagate.

The presence of a non-null AHE contribution also modifies the $\alpha^{\prime\prime}$ threshold for which the characteristic lengths are defined, as illustrated in Figs.~\ref{plot-wavelength-AHE-mode-plus}, \ref{plot-propagation-length-AHE-mode-plus}, \ref{plot-wavelength-AHE-mode-minus}, \ref{plot-propagation-length-AHE-mode-minus}. This feature constitutes a characteristic signature of polaritonic propagation in bi-isotropic media with an AHE contribution. Moreover, peaks in the propagation length $L_{SPP}^{\pm}$ can arise as well at $\tilde{\alpha}$, given by \eqref{alpha-tilde-definition-1}, for $\epsilon_{1}<0$, $\epsilon_{2}>0$, and $\epsilon_{1}<-\epsilon_{2}$, but at distinct frequency regimes for each mode. This means that $L_{SPP}^{+}$ presents a maximum value (before decreasing) at $\tilde{\alpha}$ for $(b\epsilon^{\prime}_{1})/ (\epsilon_{2}^{2}+\epsilon_{1}^{\prime}\epsilon_{2}- \epsilon_{1}^{\prime 2}) < \omega \leq b/\epsilon_{2}$, whereas, for $L_{SPP}^{-}$, the maximum occur for $0<\omega < (b\epsilon^{\prime}_{1})/ (\epsilon_{2}^{2}+\epsilon_{1}^{\prime}\epsilon_{2}- \epsilon_{1}^{\prime 2})$. Thus, these peaks are not restricted to a single mode; they can arise in both modes within distinct frequency ranges, revealing that the bi-isotropic and AHE contributions can significantly modify the surface-polariton propagation under specific conditions.

It is worth emphasizing that the multiplicity of SPs reported here refers specifically to TM solutions, in contrast to previous investigations of bi-isotropic interfaces involving topological insulators \cite{Karch}, in which the magnetoelectric factor in Eq.~(\ref{MELF}) is real, $\alpha=\alpha'$. In that case, modified hybrid SP solutions (composed of TM and TE components) were obtained. By contrast, in the scenario considered in the present work, no additional TM solutions could arise for topological insulators if the magnetoelectric factor is taken to be real. Indeed, for $\alpha=\alpha'$ the refractive indices in Eq.~(\ref{isotropic-case-1-2}) recover the usual isotropic-medium result, $n_{\pm}=\sqrt{\mu\epsilon}$, thereby reducing the multiple SP solutions to the conventional ones described by Eqs.~(\ref{kxusual1}) and (\ref{kzusual1}).

	\begin{acknowledgments}
	
The authors thank FAPEMA, CNPq, and CAPES (Brazilian research agencies) for their invaluable financial support. M.M.F. is supported by CNPq/Produtividade 317048/2023-6, and CNPq/Universal/420896/2025-2. P.D.S.S. is grateful to Produtividade/FAPEMA/CNPq/12649/25, and UFMA. Furthermore, we are indebted to CAPES/Finance Code 001 and FAPEMA/POS-GRAD-04755/24.

	\end{acknowledgments}

\end{document}